\documentclass{aa}  

\usepackage{graphicx}
\usepackage{txfonts}
\usepackage{hyperref}
\hypersetup{
    colorlinks=true,
    citecolor=blue,
    urlcolor=blue,
}
\begin{document}

   \title{Fitting infrared ice spectra with genetic modelling algorithms}

   \subtitle{Presenting the \texttt{ENIIGMA} fitting tool}

   \author{W. R. M. Rocha\thanks{Current address: Laboratory for Astrophysics, Leiden Observatory, Leiden University, P.O. Box 9513, NL 2300 RA Leiden, The Netherlands.\\
             \email{rocha@strw.leidenuniv.nl}},
          G. Perotti,
          L. E. Kristensen,
          \and
          J. K. J{\o}rgensen
          }

   \institute{Niels Bohr Institute \& Centre for Star and Planet Formation, University of Copenhagen, {\O}ster Voldgade 5-7, DK-1350 Copenhagen K., Denmark\\
             }

   \date{Received xxxx; accepted yyyy}

 
  \abstract
   {A variety of laboratory ice spectra simulating different chemical environments, ice morphology as well as thermal and energetic processing are demanded to provide an accurate interpretation of the infrared spectra of protostars. To answer which combination of laboratory data best fit the observations, an automated statistically-based computational approach becomes necessary.}
   {To introduce a new approach, based on evolutionary algorithms, to search for molecules in ice mantles via spectral decomposition of infrared observational data with laboratory ice spectra.}
   {A publicly available and open-source fitting tool, called ENIIGMA (d{\bf E}compositio{\bf N} of {\bf I}nfrared {\bf I}ce features using {\bf G}enetic {\bf M}odelling {\bf A}lgorithms), is introduced. The tool has dedicated Python functions to carry out continuum determination of the protostellar spectra, silicate extraction, spectral decomposition and statistical analysis to calculate confidence intervals and quantify degeneracy. As assessment of the code, fully blind and fully sighted tests were conducted with known ice samples and constructed mixtures. Additionally, a complete analysis of the Elias~29 spectrum was performed and compared with previous results in the literature.}
   {The \texttt{ENIIGMA} fitting tool can identify the correct ice samples and their fractions in all checks with known samples tested in this paper. In the cases where Gaussian noise was added to the experimental data, more robust genetic operators and more iterations became necessary. Concerning the Elias~29 spectrum, the broad spectral range between 2.5$-$20~$\mu$m was successfully decomposed after continuum determination and silicate extraction. This analysis allowed the identification of different molecules in the ice mantle, including a tentative detection of CH$_3$CH$_2$OH.}
   {The \texttt{ENIIGMA} is a toolbox for spectroscopy analysis of infrared spectra that is well-timed with the launch of the {\it James Webb Space Telescope}. Additionally, it allows for exploring different chemical environment and irradiation field in order to correctly interpret the astronomical observations.}

   \keywords{ISM: molecules -- solid state: volatile -- Infrared: ISM -- Stars: protostars -- Astrochemistry}

\maketitle
%

\section{Introduction}
Characterizing the composition of astrophysical ice mantles in the interstellar medium (ISM) is crucial for understanding the chemical evolution of Young Stellar Objects (YSOs). With this goal, systematic studies in the infrared (IR) with ground- and space-based telescopes \citep[e.g.,][]{Chiar1995, Schutte1996,Gibb2000, Pontoppidan2003, Boogert2008, Zasowski2009, Thi2011, Oberg2011_spitzer} aided by laboratory experiments simulating diverse chemical environments and irradiation fields \citep[e.g.,][]{Gerakines1995, Schutte1993, Palumbo1998, Watanabe2002, Fuchs2009, Pilling2010ammonium, Linnartz2015, Scheltinga2018, Rachid2020, Ioppolo2021} have pushed forward the frontiers of knowledge about the composition of interstellar ices. 

The Infrared Space Observatory (ISO) allowed for the first systematic studies of ices in star-forming regions \citep[e.g.][]{Schutte1996, vanDishoeck1998}. These observations revealed the presence of H$_2$O, NH$_3$, and CH$_4$, as well as allowed to infer the synthesis of new chemical species such as CH$_3$OH and NH$_4^+$, both byproducts of heating and energetic processing by photons and cosmic rays \citep{Gibb2000, Gibb2004}. Tentative assignments of CH$_3$CHO (acetaldehyde) and HCOOH (formic acid) relying on the absorption features at 7.24~$\mu$m and 7.41~$\mu$m were made by \citet{Schutte1999}. However, the strong bands at 5.80~$\mu$m associated with these two molecules have not been unequivocally detected yet, and, therefore the presence of these molecules is still under debate.

The {\it Spitzer} Space Telescope improved in sensitivity compared to ISO allowed ice surveys toward low-mass star-forming regions, background stars and high-UV environments \citep[e.g.,][]{Bergin2005, Whittet2009, Reach2009, Boogert2013}. Systematic analysis of the 5$-$8~$\mu$m absorption complex, CH$_4$ band at 7.7~$\mu$m, CO$_2$ bending mode at 15.2~$\mu$m, as well as NH$_3$ and CH$_3$OH absorption features at 9.01~$\mu$m and 9.74~$\mu$m, were performed in \citet{Boogert2008, Oberg2008, Pontoppidan2008} and \citet{Bottinelli2010}, respectively. All used a phenomenological approach and/or analytical function decomposition to estimate environment-dependent variations, in order to avoid degenerated solutions in characterizing the ice morphology. A statistical analysis of the ice column density and abundance with respect to H$_2$O in these four studies was done by \citet{Oberg2011_spitzer}, who concluded that except in the case of CO, CO$_2$ and CH$_4$ (the most volatile species), there is no significant statistical abundance variation between low- and high-mass YSOs in the case of OCN$^-$, CH$_3$OH, and the broad residual component, C5, described in \citet{Boogert2008}. In the case of background sources, however, the abundances of CO and CO$_2$ are similar to the low-mass YSOs and higher than the high-mass YSOs by a factor of around 3.4 \citep{Boogert2013,Boogert2015}. 

Considering pathways and morphology, the analysis of {\it Spitzer} observations demonstrated that CO$_2$, NH$_3$, and CH$_4$ ice abundances do not vary significantly with respect to H$_2$O ice, which indicates co-formation \citep{Oberg2011_spitzer}. On the contrary, the same work shows that CH$_3$OH and CO ices vary by order-of-magnitude relative to water ice, thus indicating different formation pathways. NH$_4^+$ is anti-correlated with H$_2$O, likely due to the high desorption temperature compared to other volatiles \citep{Greenberg1985}. In the case of CO and CO$_2$, there is evidence in support of cosmic-ray processing, given the abundance ratio CO:CO$_2$, as well as thermal evolution, that is mainly seen by the CO$_2$ bending mode splitting at 15.2~$\mu$m \citep{Pontoppidan2008}. Attempts at assigning more complex molecules such as HCOOH, CH$_3$CHO and CH$_3$CH$_2$OH were also made by \citet{Oberg2011_spitzer}, although they are still relying on the absorption features between 7$-$8~$\mu$m.

The main disadvantage of the phenomenological approach is the limited information about ice morphology and the {\it a priori} assumption of carriers associated with the absorption bands. In addition, these methods are usually limited to a short wavelength range to address a small set of spectral components \citep[e.g. 3-5;][]{Pontoppidan2003b, Thi2006, Boogert2008}. Other approaches combining laboratory ice spectra to fit the observed spectra have been adopted in previous works \citep{Merrill1976, Gibb2004, Thi2006, Zasowski2009, Suutarinen2015}. Although the degeneracies are high in these methods, some issues in the direct comparison of the observations with the experimental data were revealed, as is the case of the absorption peak at 6~$\mu$m attributed to H$_2$O ice, found to be deeper than the features at 3~$\mu$m and 13~$\mu$m \citep{Schutte1996, Keane2001}, as well as the nature of the peaks at 3.55~$\mu$m and 6.85~$\mu$m, that cannot be only attributed to CH$_3$OH ice \citep{Dartois2001,Schutte2003, Bottinelli2010}.

In order to allow exploitation of large data-sets of laboratory ice spectra to identify and quantify molecules in ice mantles observed toward YSOs, a new computational code for spectral analysis is presented in this paper. The code uses genetic modelling algorithm \citep{Holland1975} to decompose IR spectra of protostars using a linear combination of laboratory data themselves. The use of evolutionary algorithms have successfully been used in the literature to derive the dust composition in asymptotic giant branch (AGB) stars \citep[][]{Baier2010}, as well as to derive physical properties of protostars \citep[][]{Woitke2016}. This paper is laid out as follows: Section 2 list the public databases containing ice spectra used in this work and Section~3 details the code methodology. The accuracy and performance tests of this tool are discussed in Section~4, and an application to a real source, the Class I protostar Elias~29 is shown in Section~5. A summary of the capabilities of the \texttt{ENIIGMA} fitting tool and the results of the Elias~29 spectral analysis are given in Section 6. 

\section{The \texttt{ENIIGMA} fitting tool}
The \texttt{ENIIGMA} (``dEcompositioN of Infrared Ice features using Genetic Modelling Algorithm'') fitting tool aims for the analysis of observational spectra containing ice absorption features. As a general overview, the structure of this tool is split into three major modules; the first provides Python functions to calculate the continuum spectral energy distribution (SED) of YSOs spectra and silicate feature removal. The second focuses on the unbiased spectral decomposition using a linear combination of laboratory ice spectrum. The third offers a statistical analysis of the fits such as confidence intervals and degeneracy quantification. The \texttt{ENIIGMA} fitting tool documentation and download instructions are publicly available at \url{https://eniigma-fitting-tool.readthedocs.io}.  

\begin{figure*}
   \centering
   \includegraphics[width=\hsize]{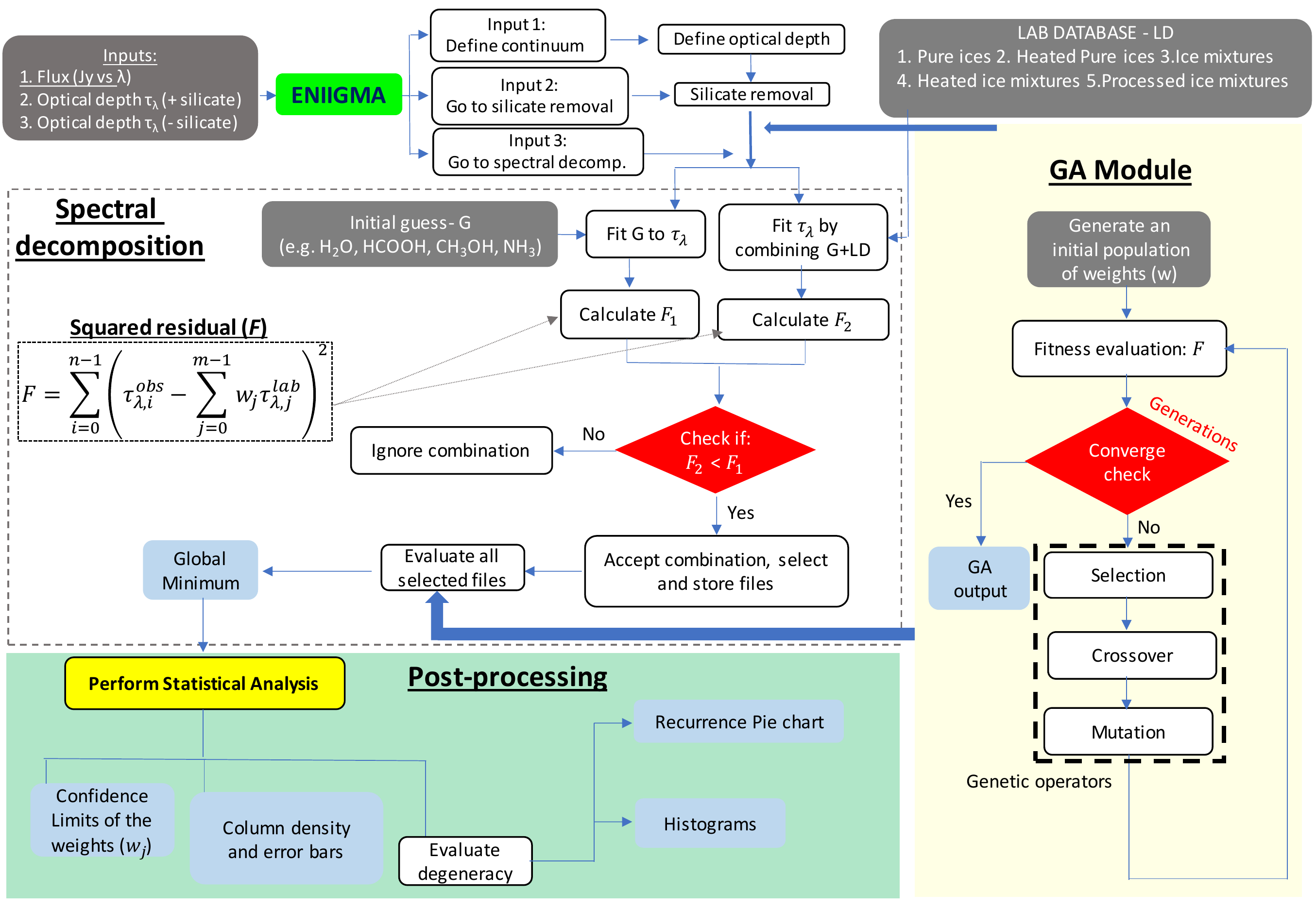}
      \caption{Flowchart of the \texttt{ENIIGMA} fitting tool. The light grey dashed box highlights the steps during the spectral decomposition. The light yellow region highlights the Genetic Algorithm (GA) module, whereas the light green region shows the steps performed during the post-processing statistical analysis. The small grey boxes indicate the input data requested by the tool. Small blue boxes are the output data. White boxes indicate the processes performed in each step.}
         \label{flowchart}
   \end{figure*}

In the first module, the \texttt{ENIIGMA} makes use of previous techniques available in the literature for the continuum determination \citep[e.g.][]{Boogert2000} and silicate removal from a template \citep[e.g.,][]{Bottinelli2010}. The second and third module stand the novelty of the code that uses genetic modelling algorithms to search for the optimal solution in a large problem space followed by a robust statistical analysis of the results. Global optimization is a mathematical method that aims to search for the best overall solution providing the maximum likelihood among the parameters. This method is unbiased in the sense that a large sample of ice spectra is tested without targeting any particular species, and the final statistically-based decision of the best fit is made. In the context of ice observations in space, it allows us to explore a significant amount of IR laboratory data and decide for the best ice components. A flowchart of the entire procedure is shown in Figure \ref{flowchart}, and the different step are detailed in the next sections.

\subsection{Input data, continuum determination and optical depth}
The \texttt{ENIIGMA} fitting tool works with three types of input data. The first possibility is to work with observed spectra representing the flux in units of Jy. The second possibility is to use spectra in an optical depth scale – including the silicate features at 9.8~$\mu$m and 18~$\mu$m. Finally, it is possible to provide the tool with spectra where the silicate features have been removed. The file format and structure must be ASCII (American Standard Code for Information Interchange), with three columns, containing the wavelength in micrometres, flux or optical depth and the errors, respectively.

The accurate analysis of the chemical composition and column densities of ices in star-forming regions depends on the subtraction of continuum SED. However, determining an accurate continuum SED of YSOs is challenging because of the large width of the ice and silicate features. Wavelengths shortwards of 5~$\mu$m are better constrained because of the lack of ice bands at the K-band and the narrow features of CO$_2$ and CO ices at 4.27 and 4.67~$\mu$m, respectively. Some methods for the continuum determination shortwards of 4~$\mu$m have been adopted in the literature such as the near-IR excess removal by using the Kurucz stellar atmosphere models \citep[e.g.][]{Pontoppidan2004} and a linear combination of black-bodies to fit near- and mid-IR photometric data \citep[e.g.][]{Ishii2002, Moultaka2004, Perotti2020, Perotti2021}. Both methods not only provide the flux continuum, but also physical properties such as near-IR extinction or effective temperature. 

The continuum above 5~$\mu$m, on the other hand, is much less constrained even though short-wavelength data is available. The entire range between 5$-$30~$\mu$m is dominated by multiple broad and narrow absorption and emission features associated to ices and polycyclic aromatic hydrocarbons (PAHs), respectively \citep[see reviews by][]{Tielens2008, Boogert2015}. Hydrogenated amorphous carbon (HAC), or more generally a-C(:H), is another class of material observed toward Galactic Center sources with multiple features in this spectral range \citep[][]{Chiar2000, Jones2012}. Toward YSOs, a-C(:H) has been attributed to the absorption bands at 3.47~$\mu$m, 6.85~$\mu$m and 7.25~$\mu$m \citep[e.g.,][]{Gibb2004, Alata2014}. Nevertheless, the profile at 3.47~$\mu$m shows a good correlation with the water ice band at 3~$\mu$m, thus suggesting this band is associated with volatile chemical species, instead \citep[e.g.,][]{Brooke1999, Dartois2002, Boogert2015}. The strong feature at 6.85~$\mu$m observed toward YSOs, has also been excluded as a potential carrier of a-C(:H)  - as it is observed in a similar wavelength in the diffuse ISM - due to the absence of the strong feature at 3.40~$\mu$m \citep[][]{Boogert2015}. The 6.85~$\mu$m band has also been attributed to the CO$_3^-$ ion in minerals, but the lack of absorption bands at short or long wavelengths discard this possibility \citep[e.g.][]{Tielens1987,Schutte1996, Boogert2008, Boogert2015}. Finally, the feature at 7.25~$\mu$m can be associated with HCOOH \citep[][]{Schutte1999}, ethanol \citep[CH$_3$CH$_2$OH;][]{Oberg2011_spitzer, Scheltinga2018} and aminomethanol \citep[NH$_2$CH$_2$OH;][]{Bossa2009}. However, \citet{Jones2016} discuss that the other compounds containing C$=$O bonds (e.g., ketone, organic carbonate), could be present in the grains before or at the onset of ice mantle formation, and therefore, the bands at 3.47~$\mu$m, 6.85~$\mu$m and 7.25~$\mu$m, could be consistent with carbonyl functional groups. Because of all of these uncertainties, a low-order polynomial fit has usually been used to trace the continuum between the 5 and 32~$\mu$m spectral range \citep[e.g.][]{Gibb2000, Gibb2004, Pontoppidan2005, Boogert2008, Zasowski2009, Boogert2011, Oberg2011_spitzer, Noble2013, Noble2017}.

Given that context, the low-order polynomial and blackbody combination methods are incorporated in the \texttt{ENIIGMA} tool for the continuum determination of the observational flux. In the polynomial approach, the function below is used:
\begin{equation}
    \label{Cont_poly}
    F_{\mathrm{poly}}(\lambda) = \sum_{k=0}^n a_k \lambda^k,
\end{equation}
where $a_k$ are the constants, $\lambda^k$ the wavelength and $k$ the polynomial order. The second method adopts a sum of blackbodies, given by:
\begin{equation}
    \label{Cont_BB}
    F_{\mathrm{BB}}(\lambda, T) = \sum_{i=1}^{m} f_i\frac{2hc^2}{\lambda^5} \left(\mathrm{exp}\left[\frac{hc}{\lambda k_\mathrm B T_i}\right] -1\right)^{-1}
\end{equation}
where $f_i$ is a scale factor, $h$ the Planck's constant, $c$ the light velocity, $k_B$ the Boltzmann's constant, $\lambda$ the wavelength and $T_i$ the temperature of each blackbody.

Once the continuum SED is determined from Equations~\ref{Cont_poly} or \ref{Cont_BB}, the observed spectrum ($F_{\lambda}^{obs}$) is converted to an optical depth scale using the following equation:
\begin{equation}
    \tau_{\lambda}^{obs} = -\mathrm{ln} \left( \frac{F_{\lambda}^{obs}}{F_{\lambda}^{cont}} \right)
    \label{tau_obs_eq}
\end{equation}

\subsection{Silicate removal}
\label{silc_removal}
The spectral range between 8$-$25~$\mu$m is dominated by strong and weaker silicate absorption features at 9.7~$\mu$m and 18~$\mu$m, respectively. However, other functional groups associated with simple or complex molecules are also blended in the silicate band at 9.7~$\mu$m. Some examples are ammonia (9.01~$\mu$m; $\nu_2$ umbrella vibrational mode) and methanol (9.74~$\mu$m; $\nu_4$ stretching vibrational mode of C$-$O bond) as discussed in \citet{Boogert2008} and \citet{Bottinelli2010}. Additionally, ethanol (CH$_3$CH$_2$OH) also shares the C$-$O stretching mode at 9.51~$\mu$m with the silicate broad profile.

A proper removal of the silicate feature is therefore critical, although not straightforward. Different methodologies have been used in the literature, such as the local continuum and silicate spectrum template of sources toward the Galactic Center \citep[e.g., the GCS~3 silicate absorption profile, taken from][]{Kemper2004}. In \citet{Boogert2008}, this absorption feature is narrower compared to the observed silicate band of  HH~46~IRS. As a consequence, when the GCS~3 absorption band is used for the silicate removal, a significant residual is left between 8$-$10~$\mu$m. A systematic study by \citet{van_Breemen2011} shows that such an IR excess is absent in the sightline of the diffuse interstellar medium, while it is present toward molecular clouds. Although this difference can be associated with ice growth in some cases \citep[e.g., H$_2$O, NH$_3$ and CH$_3$OH;][]{Ossenkopf1994, Boogert2004, Boogert2008, McClure2009, Bottinelli2010}, it is also observed toward sources with weak or absent absorption of these molecules at other wavelengths \citep[e.g., SSTc2d\_J182835.8+002616;][]{van_Breemen2011}. In this regard, such an IR excess is likely due to the chemical composition of the dust, most specifically, the combination of magnesium-rich pyroxene-type (MgSiO$_3$) and amorphous olivine-type (Mg$_2$SiO$_4$) silicates. Further evidence that the silicate band is not associated with only one composition has been found by \citet{Poteet2015}, that fitted the silicate profile with a linear combination of amorphous silicates, in particular Mg$_2$SiO$_4$, MgFeSiO$_4$, and MgSiO$_3$. This study indicates that although small, variations in the silicate band shape cannot be ruled out. In the same framework, the THEMIS (The Heterogeneous dust Evolution Model for Interstellar Solids) code also uses amorphous olivine-type and pyroxene-type silicates with inclusions of iron compounds and a-C(:H) to interpret the observed absorption bands in the diffuse ISM \citep{Jones2013}. Additionally, the effects of dust evolution from the diffuse ISM to denser regions can be addressed with the THEMIS code \citep[e.g.,][]{Kohler2015, Jones2017}.

The debate about the nature (geometry, size and composition) of the silicate feature toward galactic center sources and protostars is still to be clarified by high-quality data that will be provided by the upcoming {\it James Webb Space Telescope} observations. To avoid the uncertainties underlying the silicate feature, the \texttt{ENIIGMA} fitting tool assumes a synthetic silicate profile. This method is similar to the silicate template scaling approach used by \citet{Bottinelli2010}. The steps adopted here are the following: (I) the observational GCS~3 silicate profile is scaled to match the bands at 9.7~$\mu$m and 18~$\mu$m \citep[same as][]{Bottinelli2010}; (II) these two silicate bands are decomposed into six Gaussian components (three for each band); (III) a small variation of the width and height of the six Gaussian components are performed to match the YSO silicate profile. The synthetic silicate is therefore given by the following the set of equations:
\begin{subequations}
\begin{eqnarray}
  G(\lambda; A, \lambda_0, \sigma) &=& \frac{A}{\sigma \sqrt{2\pi}} \mathrm{exp}\left[-\frac{(\lambda - \lambda_0)^2}{2\sigma^2}\right] \\
  \tau_{\lambda}^{ss} &=& \sum_{\substack{i=0 \\ j=0}}^{\substack{m-1 \\ n-1}} G(\lambda^i, \lambda_0^j)\\ \lambda_0 (\mu m) &=& 8.3, 9.7, 11.2, 16.2, 18.0, 20.8
\end{eqnarray}
\end{subequations}
where $G(\lambda; A, \lambda_0, \sigma)$ is the Gaussian function, $\tau_{\lambda}^{ss}$ is the synthetic silicate (ss) optical depth, $\lambda_0$ are the fixed peak positions of each component, $A$ is the amplitude and $\sigma$ is the width of the components. Note that in this approach only $A$ and $\sigma$ are allowed to vary. This method assumes that the silicate width of protostars is substantially caused by the dust chemical composition as shown by \citet{van_Breemen2011} and the ice features are not affected. 

\begin{figure}
   \centering
   \includegraphics[width=\hsize]{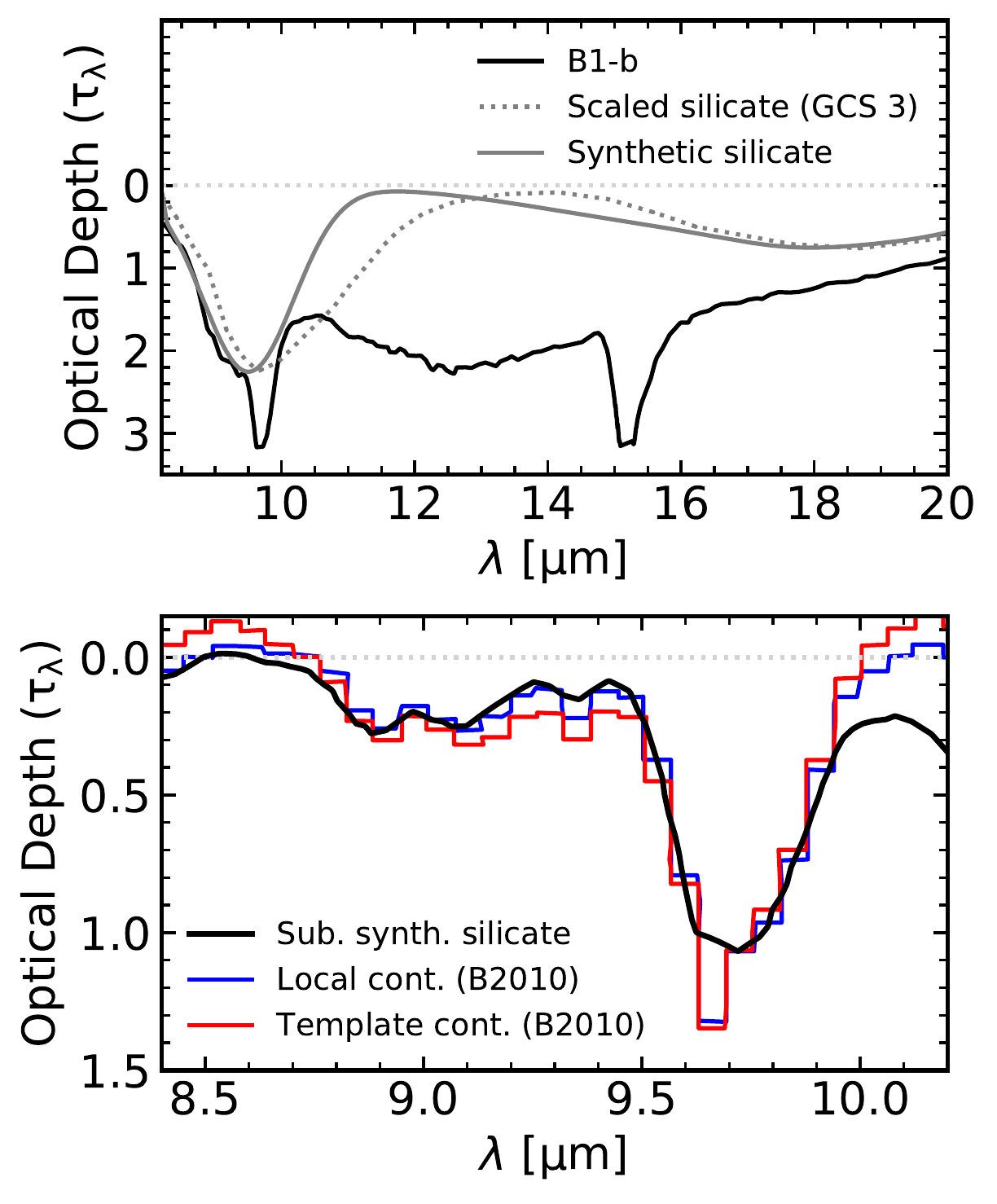}
      \caption{Illustration of the silicate removal with the synthetic silicate method. {\it Top:} The B1-b optical depth is shown by the black solid line, whereas the dashed and solid lines are the scaled GCS~3 silicate profile and the synthetic silicate feature, respectively. {\it Bottom:} Residual optical depth after synthetic silicate removal (black line). The blue and red colours show the residual optical depth after local and template continuum methods, respectively \citep[See text for details;][]{Bottinelli2010}.}
         \label{silic_B1b}
   \end{figure} 

Figure~\ref{silic_B1b} shows an example of how the synthetic silicate method is applied to an observational spectrum. As shown in the top panel, the GCS~3 silicate profile is scaled to the peak at 9.7~$\mu$m of the B1-b spectrum taken from \citet{Boogert2008} (Step I). After steps (II) and (III), the synthetic silicate feature is calculated to match the observational spectrum. The band at 18~$\mu$m, however, is less deep than the observed profile. In the bottom panel, the residual spectrum of B1-b after the silicate removal is shown and compared with the residual spectra taken from \citet{Bottinelli2010}. To extract the 9.7~$\mu$m silicate feature, the authors used the local and template continuum methods. Briefly, the local continuum mimics the silicate feature by a short-range polynomial fit, in particular, the intervals between 8.25$-$8.75, 9.23$-$9.37 and 9.98$-$10.4~$\mu$m. Conversely, the template continuum uses empirical profiles separated into three general categories, namely, (1) sources with a straight 8~$\mu$m wing, (2) sources with a curved 8~$\mu$m wing, and (3) sources with a rising 8~$\mu$m wing (``emission'' sources). In the case of B1-b source, the curved 8~$\mu$m wing has been used. As mentioned by \citet{Bottinelli2010}, the template method is not suitable to fit the 9.7~$\mu$m silicate feature toward all YSOs in their sample, and the local continuum approach was adopted instead. However, this method provides narrow bands for the feature at 9~$\mu$m, often associated with NH$_3$ ice, then in the template method \citep{Oberg2011_spitzer}. From an eyeball comparison, one can note that the synthetic silicate method introduced in this paper agrees very well with the methods adopted by \citet{Bottinelli2010}. In this way, it can be an alternative to the local continuum method when the template profile is not available. It is worth noting than an advantage of the synthetic silicate method is that it provides the silicate extraction of both  9.7~$\mu$m and 18~$\mu$m bands simultaneously, which allows a better analysis of the H$_2$O libration band at around 13~$\mu$m. Despite the challenges, new laboratory data combining ice and dust could provide a better understanding of the link between ice and dust in star-forming regions. With this goal, \citet{Potapov2018, Potapov2021} have measured spectrum and optical constants of silicate (MgSiO$_3$) and H$_2$O ice, and showed that this experimental data can partially fit the 3~$\mu$m of some astronomical objects.

\subsection{Laboratory ice spectra and databases}
\label{Lab_dat}
In order to build an internal database of experimental data for the \texttt{ENIIGMA} to decompose the spectra of YSOs, a data-set of around 100 laboratory ice spectra were compiled from publicly online databases which are listed below:
\begin{itemize}
    \item NASA Ames: The Astrophysics \& Astrochemistry Lab at NASA Ames Research Center\footnote{\url{http://www.astrochem.org}}
    \item Leiden DB: Leiden Ice Database\footnote{\url{https://icedb.strw.leidenuniv.nl}}
    \item UNIVAP: Laboratorio de Astroquimica \& Astrobiologia da UNIVAP\footnote{\url{https://www1.univap.br/gaa/nkabs-database/data.htm}}
\end{itemize}

The ice samples compiled in this paper were recorded in transmittance ($T$) or absorbance ($A$) mode using Fourier Transform Infrared Spectrometer (FTIR). All spectra in transmittance scale were converted to absorbance using the equation $A = -\mathrm{log_{10}}(T)$. In the absorbance scale, the ice spectra can be converted to optical depth using following the equation \citep{dHendecourt1986}:
\begin{equation}
    \tau_{\lambda}^{lab} = 2.3 \cdot A_{\lambda}.
    \label{od_eq}
\end{equation}

In addition to the compiled data, two IR spectra containing organic residue material (henceforth: residue) at high temperature (> 100~K) were taken from the literature. Residue is produced with or without UV irradiation of the ice mixture samples and subsequent warm-up until room temperature (300~K). In particular, the spectra taken from \citet{Schutte2003} are HNCO:NH$_3$ heated until 120~K without energetic processing and H$_2$O:CO$_2$:NH$_3$:O$_2$ irradiated with UV and heated until 200~K. In common, these experiments identified the ammonium ion (NH$_4^+$) formation, a likely carrier of the absorption band at 6.85~$\mu$m in YSOs spectra \citep[][]{Schutte1996}.

Accurate baseline correction of the laboratory data is required to avoid misinterpretations of the carriers. As an example, the band at 5.85~$\mu$m that has been attributed to HCOOH by \citet{Schutte1996, Schutte1999}, was put into question by \citet{Boogert2008} and \citet{Oberg2011_spitzer} because of unreliable laboratory baselines. Instead, ethanol (CH$_3$CH$_2$OH) was proposed as the carrier by those authors. To avoid this issue, the baselines of the data compiled in this paper were checked and corrected when necessary. Appendix~\ref{Laboratory_data_list} lists the data-sets used in this paper and the reference from where the data were taken. In brief, these ice data are grouped by common characteristics, such as (i) pure, (ii) pure and heated, (iii) heated mixtures, (iv) irradiated mixtures and (v) residue.

\subsection{Genetic Algorithm Module}
Inspired by the theory of natural selection and biological evolution, the genetic algorithms (GA) aim to provide a global optimization process for a proposed problem. For clarification purposes, the GA nomenclature is briefly explained: (i) {\it gene} corresponds to the value of a parameter (e.g., scaling factor of the ice spectrum to match the observation), (ii) {\it chromosome} is a set of values proposed as a solution of the parameters, (iii) {\it fitness function} is the criteria to evaluate the fit (e.g., squared residual, reduced $\chi^2$), {\it selection} is the election of the best parameters out of the entire set of potential solutions (chromosome) used to generate improved values, (iv) {\it crossover} is the mixing of values from two distinct solutions in order to create a new solution, (v) {\it mutation} is a small variation of one or more parameters (gene), (vi) {\it global minimum} is the optimal solution of the parameters and (vii) {\it generation} corresponds to an interaction where potential solutions are tested. 

The GA module implemented in the \texttt{ENIIGMA} fitting tool is build on Pyevolve\footnote{\url{http://pyevolve.sourceforge.net/0_6rc1}} \citep[][]{Perone2009}, an open-source and extensible library dedicated to perform evolutionary computation in Python programming language. The GA computation is composed of three main characteristics, namely, generation of random population of probable solutions, fitness-oriented to evaluate the population, and variation-driven to improve the next population \citep{Holland1975, Koza1992}. The population follows the chromosome-like structure, in which the vector $w_{ij} \in \mathbb{R}^{m \times n}$ is given by:
\begin{equation}
    \Vec{W} = \begin{bmatrix} 
    w_{11} & w_{12} & \dots & w_{1n}\\
    w_{21} & w_{22} & \dots & w_{2n}\\
    \vdots & \vdots & \ddots & \vdots\\
    w_{m1} & w_{m2} & \dots & w_{mn} 
    \end{bmatrix}
    \label{chromo_eq}
\end{equation}
where each variable $w$ is called gene, and the rows are the combination of genes that provide a solution. The quality of population is evaluated by a fitness function ($F$) - the squared residual as shown below: 
\begin{equation}
    F = \sum_{i=0}^{n-1} \left(\tau_{\lambda,i}^{obs}  - \sum_{j=0}^{m-1} w_j \tau_{\lambda,j}^{lab} \right)^2
    \label{residual_eq}
\end{equation}
where the optical depth of the observational spectrum ($\mathrm{\tau_{\lambda,i}^{obs}}$) was defined in the Equation~\ref{tau_obs_eq}, $\mathrm{\tau_{\lambda,j}^{lab}}$ is the optical depth of the laboratory data, $w_j$ is the scale factor (the genes in the chromosome; Equation~\ref{chromo_eq}), and $\mathrm{\sigma_i^2}$ is the error of optical depth, propagated from the error in the flux. 

After checking the random population with the fitness function in the first iteration, the genetic operators are used to improve the population in the subsequent generations. Selection is the first operator and aims to guarantee the survival of the best genes (coefficients) from one generation to another. One of the selection methods adopted in this paper is called ``roulette wheel'' and elects the best individuals based on their probability of providing good solutions, defined as:
\begin{equation}
    P_i = \frac{F_i}{\Sigma_{j=1}^{N} F_j}
\end{equation}
where $F_i$ is the fitness of individual $i$ in the population, and $N$ is the number of individuals in the population. The second selection method is called ``tournament'', and randomly selects a fixed number of individuals from the entire population. The ``winner'' of the tournament, i.e., the coefficient with the highest fitness is passed to the next generations. In comparison with the roulette wheel method, the tournament selection has a lower computational cost. Moreover, the roulette wheel method suffers from the problem of premature convergence if there are individuals with high probability to be selected, whereas the selection pressure in the tournament selection is constant and keep the diversity of the population \citep{Tobias1996}. 

The best individuals passed from a generation to another are elected to crossover operation. Formally, if Equation~\ref{chromo_eq} is given for only two chromosome of two genes, i.e., $\Vec{W^1} = [w_{11}, w_{12}]$ and $\Vec{W^2} = [w_{21}, w_{22}]$, the single-point crossover operator is given by $H_{W^1,W^2}$, which will allow the elected genes to exchange their position among each other. Depending on the complexity of the problem, for example, a linear combination of more than four components, two-point or multi-point cross-overs can also be used in the \texttt{ENIIGMA}. The crossover rate is defined as 80\% in Pyevolve and provided good solutions for many problems in the literature \citep[e.g.,][]{Peng2003, Baier2010}. However, a crossover rate of 90\% provided better results in this paper when the tournament selection method was adopted. The mutation operator is used next, and consists of applying a small perturbation $\zeta$ in one or more genes (scale factor), formally expressed as $w_{mn}^{'} = w_{mn} + \zeta$. In particular, a Gaussian mutation is adopted with a mutation rate of 10\% in this paper. Compared with other mutation methods, a Gaussian distribution of the genes has found to be superior to other approaches \citep[][]{Hinterding1995}. Once the genetic operators are applied, the fitness function evaluates how close to the solution the new genes are. This process is repeated until the convergence criteria are reached, that in this paper, is the number of generations.

Although it is one of the simplest random-based Evolutionary Algorithms (EAs), it does not mean  that it is less robust than other EAs or even other conventional optimization methods. As such, it has been used to solve complex problems in astrophysics such as the huge degeneracy behind the physical parameters of protoplanetary disks and the silicate composition in the mid-IR \citep[e.g.][]{Charbonneau1995, Hetem2007, Baier2010, Woitke2019}.

\subsection{Searching for the best solution}
The \texttt{ENIIGMA} fitting tool requires an initial guess ($G$) composed of four laboratory spectra to start searching for the best combination that minimizes the residual, in Equation~\ref{residual_eq}. At this initial stage, the GA module is applied to fit $G$ to the observational optical depth $\tau_{\lambda}^{obs}$, and the residual $F{_1}$ is calculated. Next, each laboratory data is added to the initial guess one at a time in a numerical loop, and another combination is used to fit $\tau_{\lambda}^{obs}$, that leads to another residual calculation, $F{_2}$. All the laboratory data providing  $F{_2} < F{_1}$ are stored, and after this stage, the GA module is applied to combine them in groups of 8 or more genes, which results in the global minimum. It is worth noting that during the search for the best solution, the information from the initial guess might be lost since other ice spectra can provide better solutions. In this sense, the \texttt{ENIIGMA} fitting tool does not target a specific molecule or ice sample, but rather it searches for the best linear combination from a large database performing an unbiased search for the global minimum solution.

The population ($\Vec{W}$) and the generation numbers are two of the most important parameters in GAs. If these two numbers are too small, the GA module cannot find a proper solution, since the variability of the genes is essential in the context of EAs. High numbers, otherwise, will provide good solutions, but the computation limit must be also taken into account. Following previous works employing GAs \citep[e.g.,][]{Szkody2010, Harrison2015}, the \texttt{ENIIGMA} fitting tool uses the population-to-generation ratio of 90/100 or 150/200, which allows it to search the global minimum parameters from a large population avoiding premature convergence.

\subsection{Ice column density}
The optical depths obtained from the GA optimization are used to calculate the ice column densities, following the equation:
\begin{equation}
    N_{ice} = \frac{1}{\mathcal{A}} \int_{\nu_1}^{\nu_2} (w \cdot \tau_{\nu}) d\nu
    \label{CD_eq}
\end{equation}
where $\mathcal{A}$ is the band strength of a specific vibrational mode. If the ice is composed of only one molecule (pure ice), the absorption features themselves are used to calculate the column density. In the opposite case, however, blended features are avoided, and the isolated vibrational modes are used instead. If, however, blended profiles cannot be avoided, a Gaussian or Lorentzian decomposition is applied, aiming to separate the components. Table~\ref{ice_bs} lists the band strengths of molecules present in the \texttt{ENIIGMA} database used to calculate the ice column densities.

\begin{table*}
\caption{\label{ice_bs} Ice features and band strengths of the identified molecules during the spectral decomposition.}
\centering 
\begin{tabular}{lccccc}
\hline\hline
Molecule & $\lambda \; [\mu \mathrm{m}]$ & $\nu \; \mathrm{[cm^{-1}]}$ & Identification & $\mathcal{A} \; \mathrm{[cm \; molec^{-1}]}$ & References\\
\hline
H$_2$O          & 3.01    & 3,322 & O$-$H stretch & $\mathrm{2.2 \times 10^{-16}}$ & \citet{Bouilloud2015}\\
H$_2$O          & 6.00    & 1,666 & H$_2$O bend & $\mathrm{1.1 \times 10^{-17}}$ & \citet{Bouilloud2015} \\
NH$_3$          & 9.01   & 1,109 & NH$_3$ umbrella & $\mathrm{2.1 \times 10^{-17}}$ & \citet{Bouilloud2015}\\
CH$_4$          & 7.67    & 1,303 & CH$_4$ deformation & $\mathrm{8.4 \times 10^{-18}}$ & \citet{Bouilloud2015}\\
CH$_3$OH        & 9.74    & 1,128 & C$-$O stretch & $\mathrm{1.78 \times 10^{-17}}$ & \citet{Bouilloud2015}\\
CH$_3$CN        & 4.44    & 2,252 & CN stretch & $\mathrm{2.2 \times 10^{-18}}$\tablefootmark{$^{a}$} & \citet{Hudson2004}\\
HCOOH           & 8.22    & 1,216 & C$-$O stretch     & $\mathrm{8.22 \times 10^{-17}}$ & \citet{Bouilloud2015}\\
H$_2$CO         & 5.83    & 1,715 & C$-$O stretch & $\mathrm{9.60 \times 10^{-18}}$ & \citet{Schutte1993}\\
CH$_3$CH$_2$OH  & 9.17    & 1,090 &  CH$_3$ rock   & $\mathrm{7.35 \times 10^{-18}}$ & \citet{Scheltinga2018}\\
CH$_3$OCH$_3$   & 8.59    & 1,163 & COC stretch. + CH$_3$ rock     & $\mathrm{3.00 \times 10^{-17}}$ & \citet{Scheltinga2018}\\
CH$_3$CHO       & 5.80    & 1,723 &  C$-$O stretch  & $\mathrm{1.3 \times 10^{-17}}$ & \citet{Scheltinga2018}\\
COO$^-$         & 6.30    & 1,586    & COO$^-$ stretch  & $\mathrm{6.0 \times 10^{-17}}$ & \citet{Caro2003}\\
NH$_4^+$        & 6.85    & 1,460      & NH$_4^+$ bend & $\mathrm{3.0 \times 10^{-17}}$ & \citet{Schutte2003}\\
CO$_2$        & 4.27    & 2,341      & CO stretch & $\mathrm{1.3 \times 10^{-16}}$ & \citet{Bouilloud2015}\\
CO        & 4.67    & 2,141      & CO stretch & $\mathrm{1.4 \times 10^{-17}}$ & \citet{Bouilloud2015}\\
\hline
\end{tabular}
\tablefoot{
\tablefoottext{$^a$}{Since we were unable to find the band strength for CH$_3$CN at 9.61~$\mu$m (CH$_3$ rock), the band at 4.44~$\mu$m (C$\equiv$N) was used.}
}
\end{table*}

\subsection{Statistical Analysis}
\label{Stats}
The heuristic nature of GA optimization makes the final result unpredictable, although one expects that the global minimum is always found. In addition to that, the decomposition of the IR ice features is a source of huge degeneracy itself on top of the fitting procedure. Given these two aspects, a statistical analysis becomes necessary to evaluate how good the solutions are after an enormous number of combinations of laboratory ice data. For example, if the GA module is applied to combine 13 ice spectra in groups of 8 and without repetition, there are 1287 different solutions for the same problem. The aim of the post-processing statistical analysis is to check how close the other combinations of the optimal solution are. In this way, both the heuristic aspect and degeneracy itself can be evaluated as a whole.

The statistical analysis is based on the minimum $\Delta\chi^2$ method \citep{Avni1980}, given by the following equations:
\begin{subequations}
\begin{eqnarray}
\chi^2 &=& \frac{F}{\sigma^2} \\
\Delta \chi^2(\nu,\alpha) &=& \chi^2 - \chi_{min}^2
\end{eqnarray}
\label{Conf_den}
\end{subequations}
where $F$ is the squared residual in Equation~\ref{residual_eq} and $\sigma$ the standard deviation taken from the observational optical depth; $\nu$ and $\alpha$ are the number of free parameters and the statistical significance, respectively. $\chi_{min}^2$ corresponds to the goodness-of-fit in the global minimum solution. One can observe, that confidence regions do not depend on the accuracy of the fit, but the number of free parameters. Equations~\ref{Conf_den}{\it a,b} are calculated from a normal variation around the optimal scale factors $w$, by using a probability distribution $p(x)$ available in the \texttt{numpy.random.normal} numpy \citep[][]{Harris2020} routine: 
\begin{equation}
    p(x) = \frac{1}{\sqrt{ 2 \pi \sigma^2 }} \mathrm{exp}\left[ - \frac{ (x - \mu)^2 } {2 \sigma^2} \right]
    \label{prob_dist_eq}
\end{equation}
where the mean, $\mu$, are the optimal scale factors themselves, and $\sigma$ the standard deviation taken from the observational optical depth. From this analysis, the confidence intervals of the linear combination coefficients are calculated with a significance of 3$\sigma$.


{\bf \underline{Column density error bars:}} The uncertainty in the column densities is calculated from the confidence intervals derived with the minimum $\chi^2$ method. In this regard, the lower and upper scale parameters are used to calculate the minimum and maximum optical depths. The ice column density calculated from these limits provides the bounds for the error bars of each molecule found during the spectral decomposition. It must be stressed that in this method the error bar depends on the number of parameters in fit instead of the covariance matrix in non-heuristic methods (e.g.,  non-linear least squares problems). For example, for a fit with two components, the 1$-$3$\sigma$ confidence intervals are, respectively obtained for $\Delta\chi^2$ equal to 2.41, 4.61 and 9.21. On the other hand, if 8 laboratory data are used, the confidence intervals become $\Delta\chi^2$ equal to 9.52, 13.36 and 20.09, respectively. 


{\bf \underline{Recurrence plots.}} The degeneracy of the spectral decomposition is quantified with recurrence plot analysis. All solutions found by the \texttt{ENIIGMA} are sorted in increasing order by the $\chi^2$ value of the fit. Next, Equation~\ref{Conf_den}b is used to calculate $\Delta\chi^2$ and define the bounds for 3$\sigma$ confidence interval. All solution outside this level of significance is excluded from this analysis. Finally, the frequency of each ice sample is calculated, and the recurrence derived by the following equation:
\begin{equation}
    R = \frac{f_i}{S}
\end{equation}
where $f_i$ is the absolute frequency of sample $i$ and $S$ is the sum of all solutions. The recurrence of the ice data is shown in percentage, and 100\% means that a given sample was found in all solutions, indicating that such a component is required for the spectral decomposition. Lower percentage value means that the corresponding ice sample can be replaced by another data and still provide a solution with high statistical significance. 


{\bf \underline{Histograms.}} Because of the degeneracy involved in the fit of observational YSO spectrum, ice column densities derived from local minimum solutions might deviate from values calculated from the optimal combination. In this regard, the column density variation of the ice components shown in the Recurrence plot is estimated with histograms analysis. The bin sizes are proportional to the column density variances and were calculated by the Freedman Diaconis Estimator taken from the \texttt{numpy.histogram$\_$bin$\_$edges} Python package. Additionally, the histograms are log-transformed and normalized by the median ice column density. With this analysis, both the mean and 3$\sigma$ confidence interval for the column densities are calculated (See also Appendix~\ref{Ap_hist}).

 \section{Accuracy tests}
In this section, we address the capability of the \texttt{ENIIGMA} fitting tool to provide accurate solutions. For these tests, known ice samples present in the database (see Appendix~\ref{Laboratory_data_list}) are used as input data to check the accuracy of the tool to search for the correct sample and fractionation.

Depending on the complexity of the problem, the method to find the global minimum solution with genetic algorithms may require more robust genetic operators. In this section, three methods were adopted to tackle the proposed problems, and their parameters are shown in Table~\ref{gen_op}. The population size and generation number are important input values. If lower values are adopted for these two parameters, the solution space might not be entirely covered. In addition, two selection methods and crossover rates were adopted, depending on the population size.

\begin{table*}
\caption{\label{gen_op} Laboratory data of pure and thermally-processed ices used in this paper}
\renewcommand{\arraystretch}{1.0}
\centering 
\begin{tabular}{lccccccc}
\hline\hline
Method & Mutation rate & Selection method & Crossover rate & Crossover method & Gen. & Pop. size\\
\hline
1 & 0.1\% & Roulette wheel & 80\% & Single-point & 100 & 90\\
2 & 0.1\% & Tournament & 90\% & Single-point & 200 & 150\\
3 & 0.1\% & Tournament & 90\% & Two-point & 200 & 150 &\\

\hline
\end{tabular}
\end{table*}

\subsection{Identification of known samples}
The ice spectra used to test the \texttt{ENIIGMA} tool with known ice samples were taken from the public databases shown in Section~\ref{Lab_dat}, and are listed in Tables~\ref{ice_list_pure} and ~\ref{ice_list_mi}.

Three tests were performed in this section, namely, (i) fully sighted: the solution is indicated in the initial guess and is also present in the database; (ii) fully blind: the solution is not included in the initial guess but it is available in the database, and (iii) solution not available: the correct sample is deliberately removed from the database and not chosen as initial guess. In all cases, the three searching methods in Table~\ref{gen_op} provided the expected global minimum solution.

\begin{figure*}
   \centering
   \includegraphics[width=\hsize]{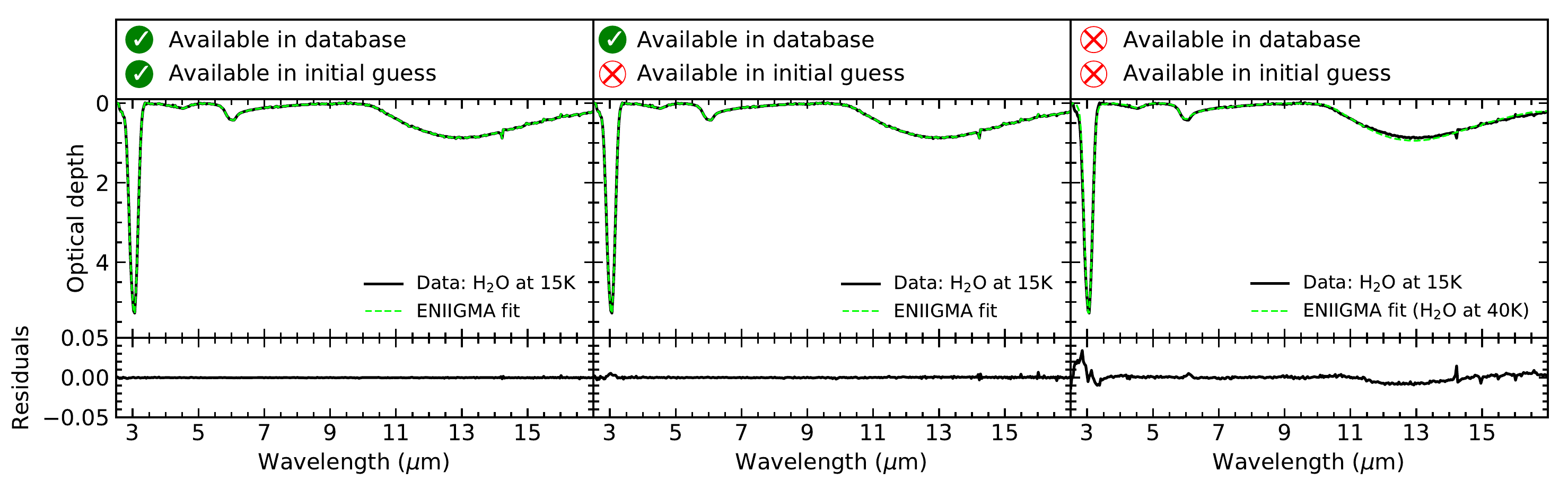}
      \caption{\texttt{ENIIGMA} fit (dashed green line) of the pure H$_2$O ice sample at 15~K (black line). Left and middle panels show the fit in fully sighted and blinded tests, respectively. The right panel shows the fit when the correct solution does not exist. The residuals of the fit are shown in the boxes below the fits.}
         \label{h2o_pure_ice}
   \end{figure*}

\begin{figure*}
   \centering
   \includegraphics[width=\hsize]{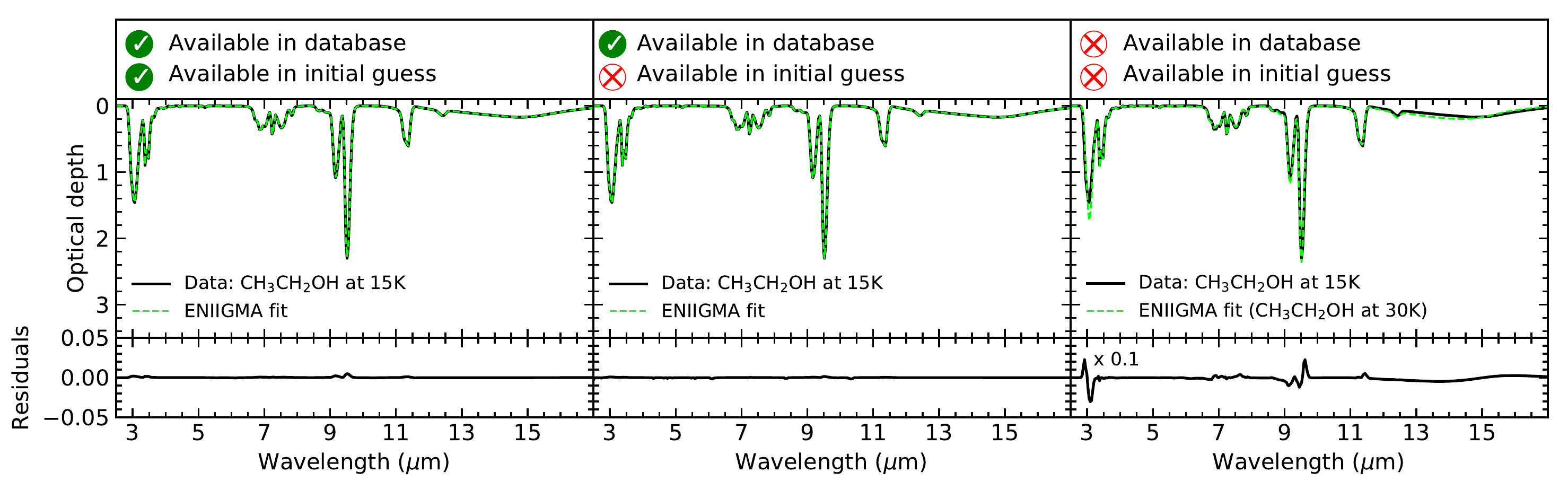}
      \caption{Same as Figure~\ref{h2o_pure_ice}, but for pure CH$_3$CH$_2$OH at 15~K.}
         \label{pure_ice}
   \end{figure*}

{\bf \underline{Pure ice:}} The test with pure ice sample is performed with H$_2$O and \citep[][]{Hudgins1993} ethanol \citep[CH$_3$CH$_2$OH;][]{Scheltinga2018} ice at 15~K. Water ice is selected because it is the most abundant molecule in the solid phase observed toward YSOs. On the other hand, ethanol was selected because of the multiple narrow features in the IR spectrum, as well as because it has some common absorption features with methanol (CH$_3$OH) ice, which is also observed in star-forming regions. Figures~\ref{h2o_pure_ice} and \ref{pure_ice} show the results of the three tests. The fully sighted and blind tests in the left and middle panels indicate that the \texttt{ENIIGMA} tool can identify the correct ice spectra associated with pure water and ethanol. The residuals shown in the bottom boxes are below 5\% and are the result of the intrinsic randomness behind the genetic optimization. It is worth noting that the difference in the residuals in the tests 1 and 2 are not associated to the presence or absence of the correct ice sample in the initial guess, but rather are related to the randomness of method itself. The right panel shows the fit when the correct solution is not available in the database. In that case, the \texttt{ENIIGMA} tool fits the input spectra with the most similar data available in the database, i.e., H$_2$O at 40~K and CH$_3$CH$_2$OH at 30~K, respectively. As a result, the residuals in some wavelengths are higher than in the tests 1 and 2.

\begin{figure*}
   \centering
   \includegraphics[width=\hsize]{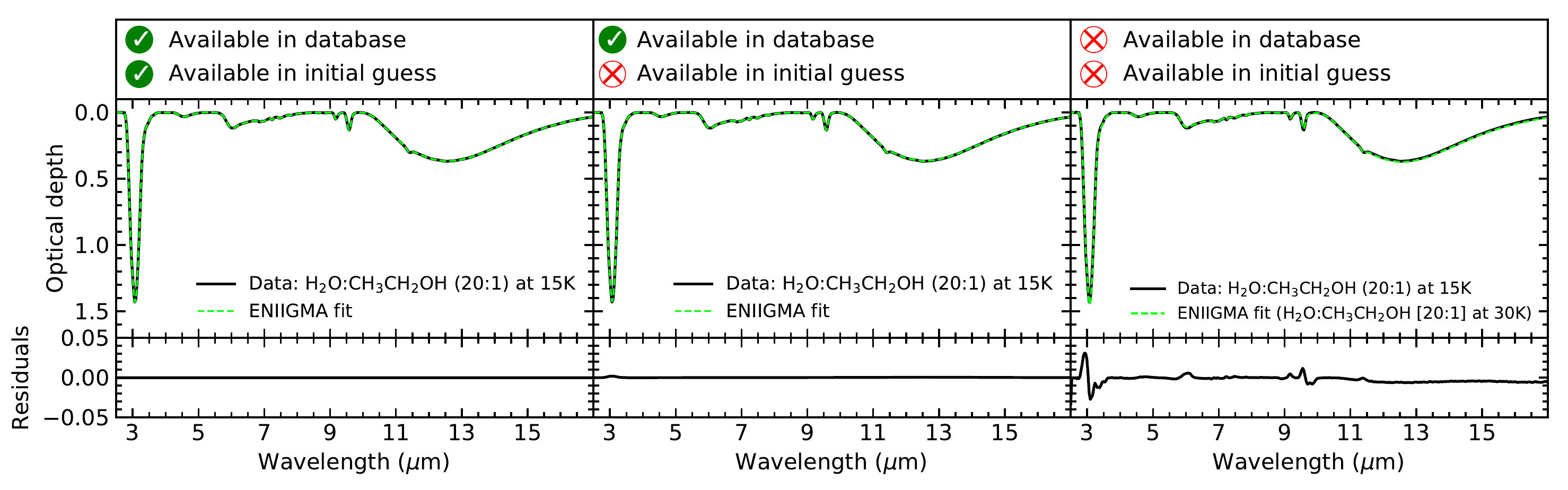}
      \caption{Same as Figure~\ref{h2o_pure_ice}, but for H$_2$O:CH$_3$CH$_2$OH (20:1) at 15~K.
              }
         \label{mix2}
   \end{figure*}

{\bf \underline{Binary ice mixture:}} A mixture dominated by H$_2$O (95\%) ice containing a fraction of ethanol (5\%) at 15~K is used in the test with binary ice sample. Despite the small fraction, ethanol shows clearly features around 9~$\mu$m and 11~$\mu$m. The fitting results are shown in Figure~\ref{mix2}. The left and middle panels showing the fully sighted and fully blinded tests indicate that the correct ice spectra were assigned with residuals lower than 5\%. In the case where the correct solution is not available (right panel), the H$_2$O:CH$_3$CH$_2$OH at 30~K was selected by the fitting routine. As in the pure ice case, this fit indicates that the \texttt{ENIIGMA} tool is able to select the most similar ice sample to fit the input spectrum. Because of the small differences of this ice mixture at 15~K and 30~K, the residual is around 5\%.   

\begin{figure*}
   \centering
   \includegraphics[width=\hsize]{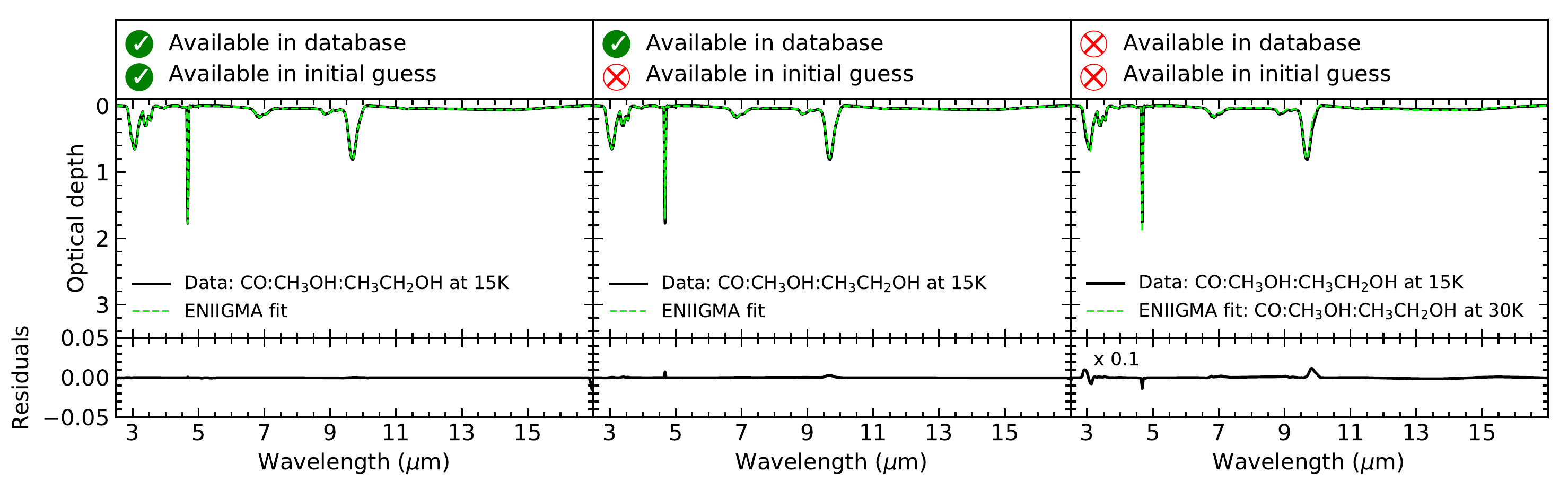}
      \caption{Same as Figure~\ref{h2o_pure_ice}, but for CO:CH$_3$OH:CH$_3$CH$_2$OH (20:20:1) at 15~K.
              }
         \label{mix3}
   \end{figure*}

{\bf \underline{Ternary ice mixture:}}
The third test is performed with a tertiary ice mixture composed by CO:CH$_3$OH:CH$_3$CH$_2$OH (20:20:1). As seen in Figure~\ref{mix3}, the fits were also accurate for the fully sighted and blinded tests (left and right panels). The residual in the middle panel is higher than in the left panel, clearly showing the random nature of the method. The fit in the right panel was performed by combining two ice samples, namely, CO:CH$_3$CH$_2$OH at 30~K and CH$_3$CH$_2$OH at 30~K. Because of the prominent CO feature in the input spectrum, not present in the ethanol ice, the code combined the two samples in order to provide a good fit. The reason why the pure CO ice was not used in this fit is that its full width half maximum is 30\% narrower than in the mixture with ethanol.

\begin{figure*}
   \centering
   \includegraphics[width=\hsize]{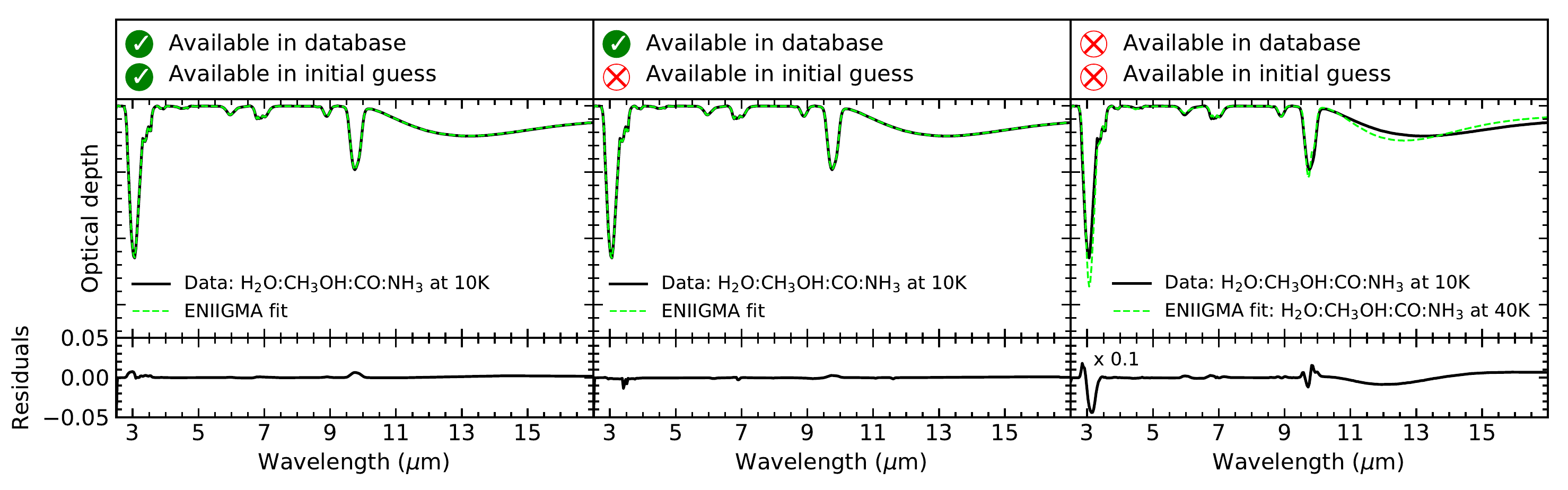}
      \caption{Same as Figure~\ref{h2o_pure_ice}, but for H$_2$O:CO:NH$_3$:CH$_3$OH at 10~K.
              }
         \label{mix4}
   \end{figure*}

{\bf \underline{Quaternary ice mixture:}} The final test about the identification of known samples is performed with H$_2$O:CH$_3$OH:CO:NH$_3$ (100:50:1:1) at 10~K, where both water and methanol features are prominent. Left and middle panels of Figure~\ref{mix4} show the result in the fully sighted and blinded tests, respectively. Again, the fits result in residuals lower than 5\%. Slightly higher residual are seen in the right panel, where the solution is not available in the database. In that case, the ice sample composed by the sample molecules at 40~K was provided with the fit. As in the case of the tertiary mixture, this result indicates that a combination of the pure molecules does not result in a good fit. For example, \citet{Dawes2016} show that the peak position of the O$-$H vibrational at around 3.0~$\mu$m is redshifted in a mixture of H$_2$O:CH$_3$OH compared to the pure H$_2$O. Nevertheless, the temperature difference of this mixture at 10 and 40~K is evident by the residuals. At this temperature, most of the CO ice has been thermal desorbed \citep{Collings2004, Acharyya2007}, whereas a small fraction trapped in the ice porous has enough energy to migrate over active sites.

\subsection{Fractionation}
As water represents the major component of astrophysical ices observed toward YSOs, this section aims to check the capability of the \texttt{ENIIGMA} tool in finding the correct coefficients in a linear combination of H$_2$O ice and another component. In the tests shown in Sections~\ref{test1}-\ref{test3}, two IR ice spectra were summed with different coefficients and Gaussian noise of standard deviation equal to 0.01 has been added to the data to resemble the noise present in the astronomical observations. Although the tests in Sections~\ref{test1} and ~\ref{test2} use water ice spectra at two different temperatures, the aim is to discuss the fractionation of constructed mixtures with similar spectral bands, rather than thermal processing. The searching Methods 1$-$3 in Table~\ref{gen_op} provided the expected global minimum solutions. In the test shown in Section~\ref{test4}, eight IR ice spectrum were considered to mimic the ice absorption features observed toward protostars. In this case, only Methods 2 and 3 provided the best fits.

\subsubsection{H$_2$O (15K) and H$_2$O (40K)}
\label{test1}
\begin{figure*}
   \centering
   \includegraphics[width=\hsize]{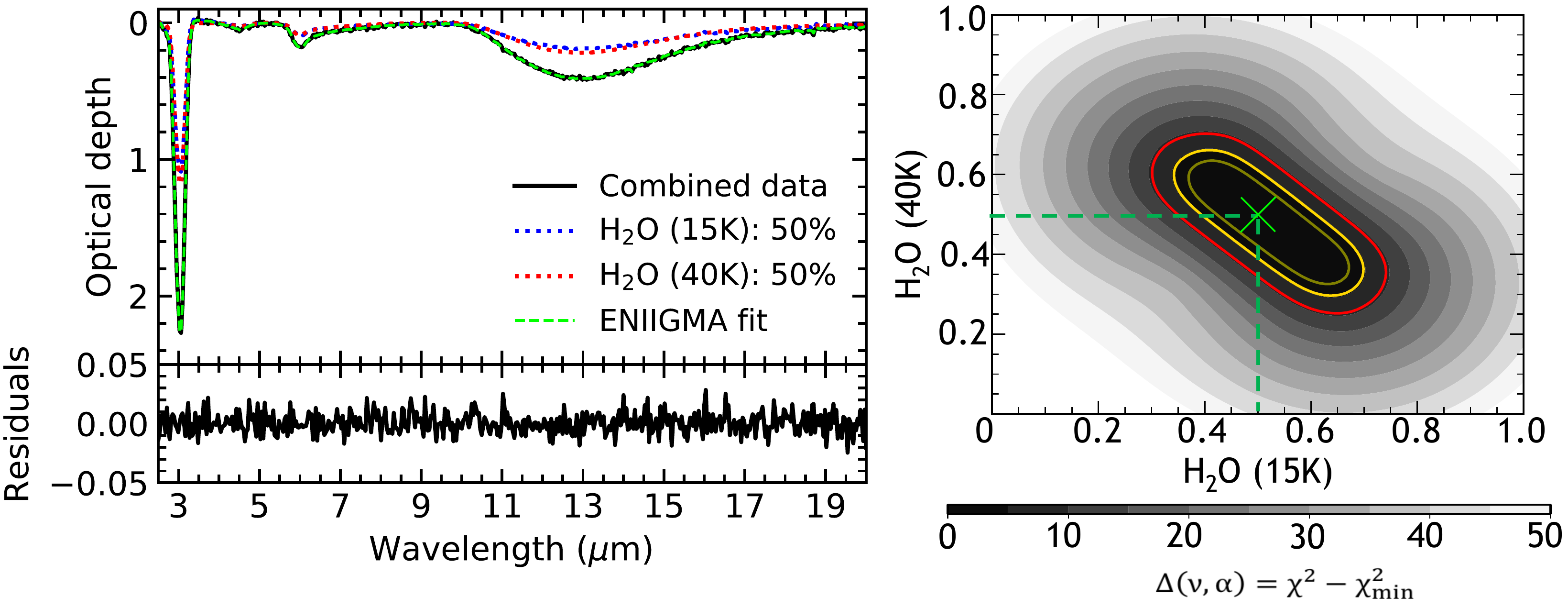}
    \includegraphics[width=\hsize]{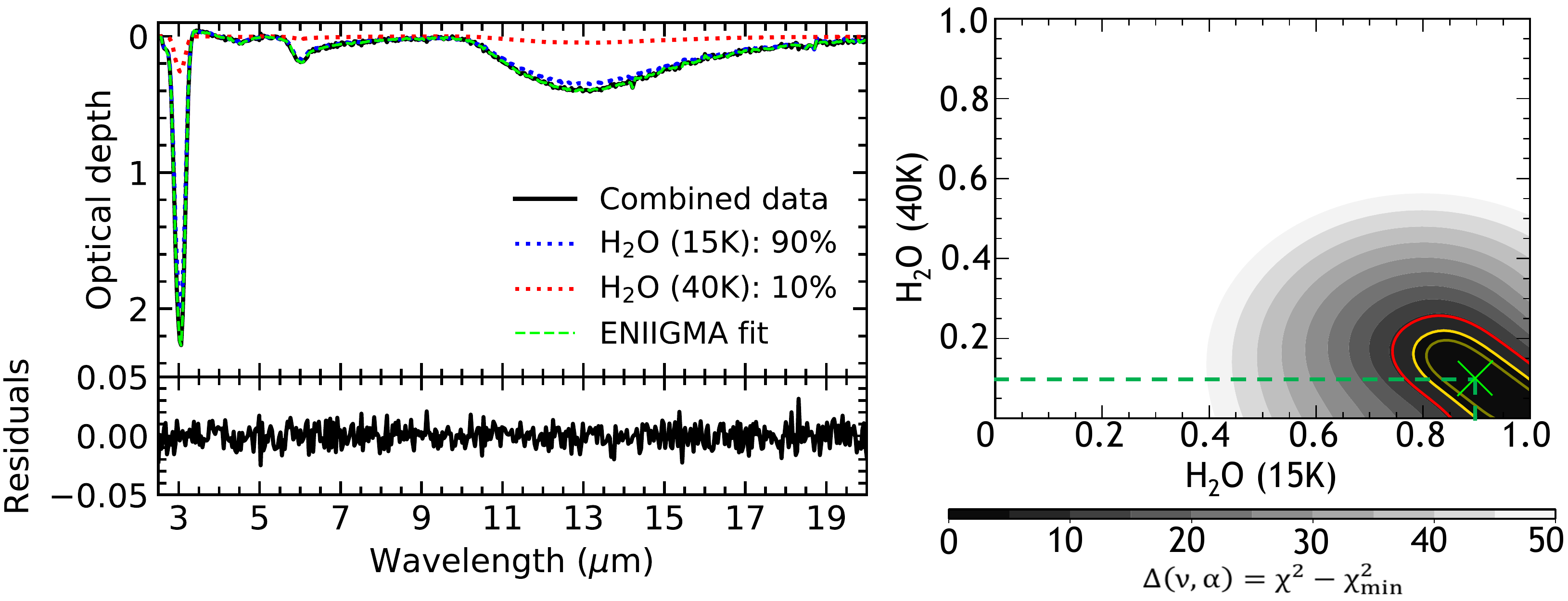}
      \caption{Spectral decomposition and confidence interval analysis of the \texttt{ENIIGMA} results. {\it Left} panels show the spectral decomposition of the combined water ice spectra (black solid line) with proportions of 50:50 and 90:10. The components of the decomposition are given by the red and blue dotted lines, whereas the \texttt{ENIIGMA} fit is given by the green dashed line. The residual of the fit is shown in the bottom parts of these graphs. {\it Right} panels show the confidence interval analysis of the coefficients in the linear combination. The greyscale colours are the difference between chi-square and minimum chi-square values; the olive, yellow and red contours are 1$\sigma$, 2$\sigma$ and 3$\sigma$ confidence intervals, respectively.}
         \label{Frac1}
   \end{figure*}

IR spectra of pure water at 15~K and 40~K were summed with proportions of 50:50 and 90:10, respectively. These two ice data were selected because they have small differences in the band shapes and thus represents a challenging task for fitting methods. Figure~\ref{Frac1} shows the test results for these two H$_2$O ice spectra. The top left panel shows the good match between the \texttt{ENIIGMA} fit and combined H$_2$O ice data with proportions of 50:50. The residual plot in the bottom part of this figure shows that only the Gaussian residual noise is left after subtracting the model from data. The top right panel shows a $\Delta \chi^2$ map, where the confidence intervals (1$-$3$\sigma$) are indicated by the green, yellow and red contours, respectively. As one can note in this degeneracy analysis, the increasing of the H$_2$O proportion at 15~K with simultaneous decreasing of the proportion of H$_2$O at (40~K) also provides statistically significant results inside the confidence intervals.

The two bottom panels of Figure~\ref{Frac1} shows the same test for the proportion of 90:10 at 15~K and 40~K, respectively. The spectral decomposition is shown in the left panel also shows a good match between the \texttt{ENIIGMA} fit and the combined H$_2$O ice data, with residuals lower than 5\%. In this test, the $\Delta \chi^2$ map shown in the right panel indicates that a solution considering only H$_2$O (15~K) is also statistically possible when the fraction of H$_2$O (40~K) is lower than 10\%. The same solution is still found by the \texttt{ENIIGMA} tool with the $\sigma_{\rm{noise}}$ equal to 0.02. Above that limit, the \texttt{ENIIGMA} is unable to find the H$_2$O (40~K) component.

\begin{figure}
   \centering
   \includegraphics[width=\hsize]{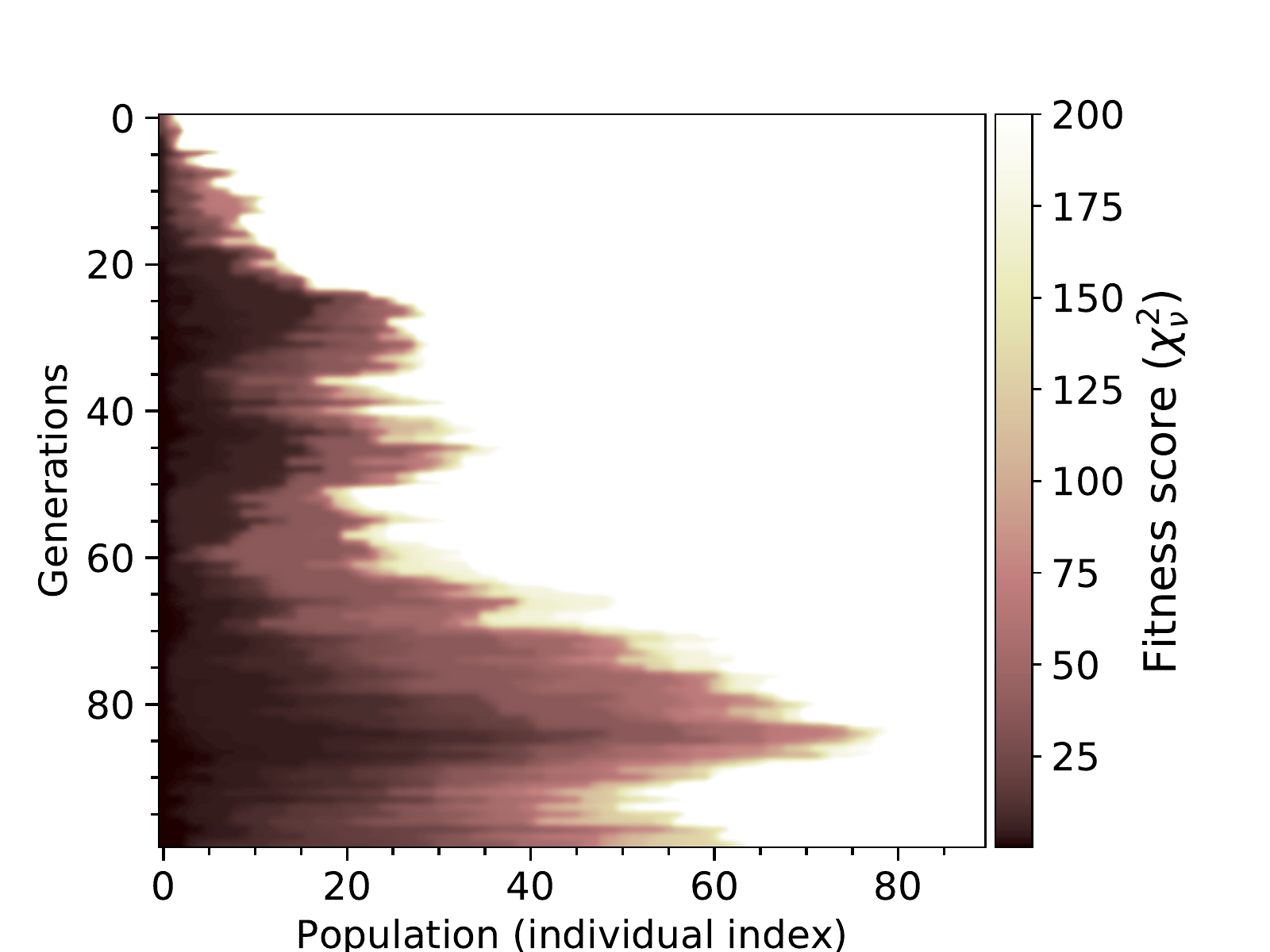}
   \includegraphics[width=\hsize]{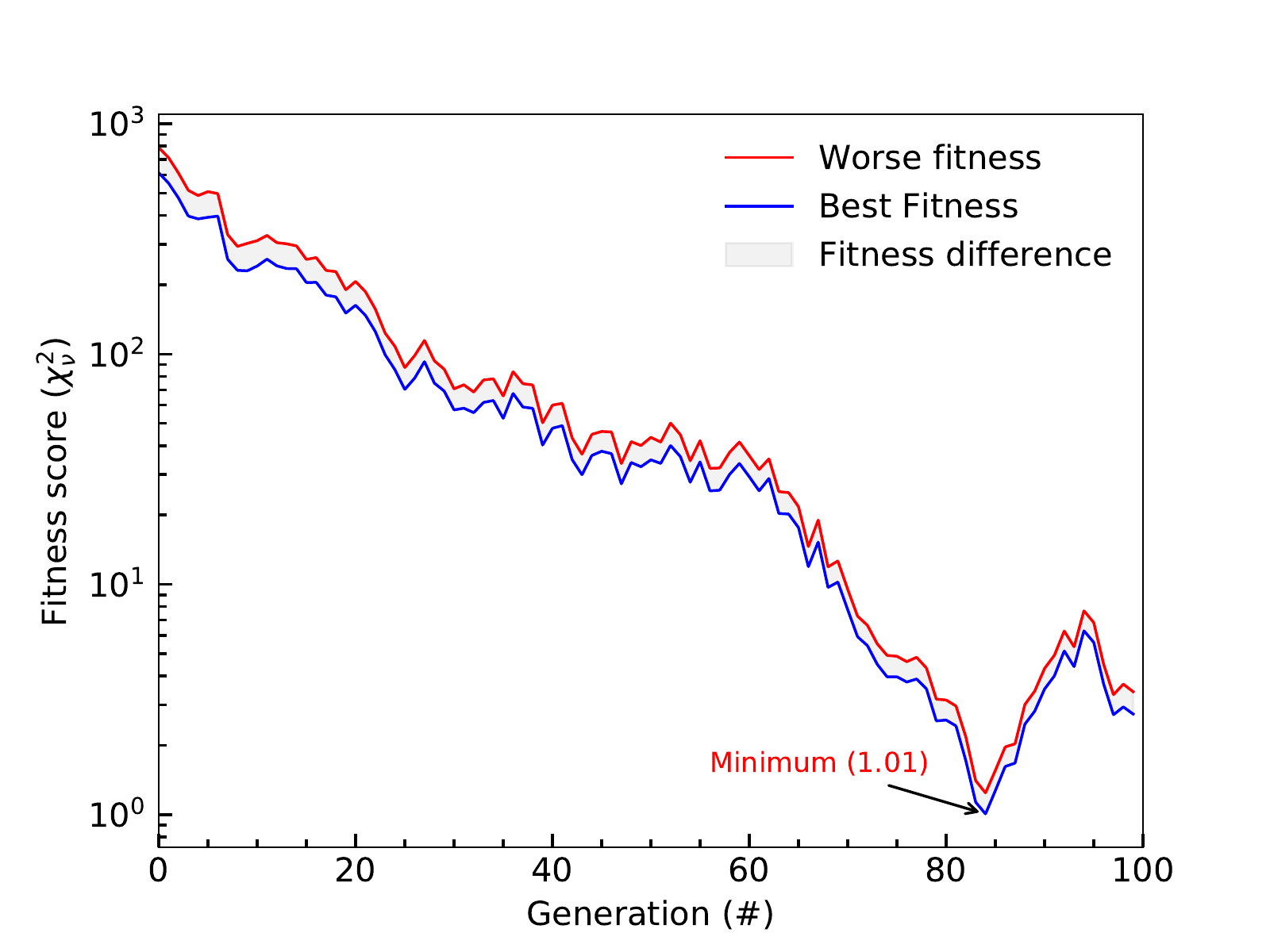}
      \caption{{\it Top:} Heat map of the fitness score for each population over the generations. {\it Bottom:} Fitness score over the generations. The blue and red colours indicate the best and worse fitness based on the population of each generation. The grey shaded area between the two lines shows the difference between the worse and best fitness. 
              }
         \label{pop}
   \end{figure}

For the case where the H$_2$O ice proportions are 90:10, a population evolution analysis is shown in Figure~\ref{pop}. At the initial generations, most of the population values result in high fitness score (white region) and therefore the optimal coefficients have not been found yet. As the optimization evolves, the population is improved and better coefficients are generated because of the survival of the best solution. Close to generation 85, most of the population results is low fitness score, which increases the probability of finding the global minimum solution of the problem. The fitness score evolution over the generations is shown in the bottom panel of Figure~\ref{pop}. As noted in the top panel, the fitness score at the initial generations is high for any value in the population. With the improvement of the population, the fit is improved until finding the optimal solution between generations 80 and 85, with $\chi_{\nu}^2 = 1.01$. As the population has not converged in this problem, the fitness score starts increasing again until generation 100, which is the stopping criteria in this problem.     
   
\subsubsection{H$_2$O (15~K) and H$_2$O (75~K)}
\label{test2}
\begin{figure*}
   \centering
   \includegraphics[width=\hsize]{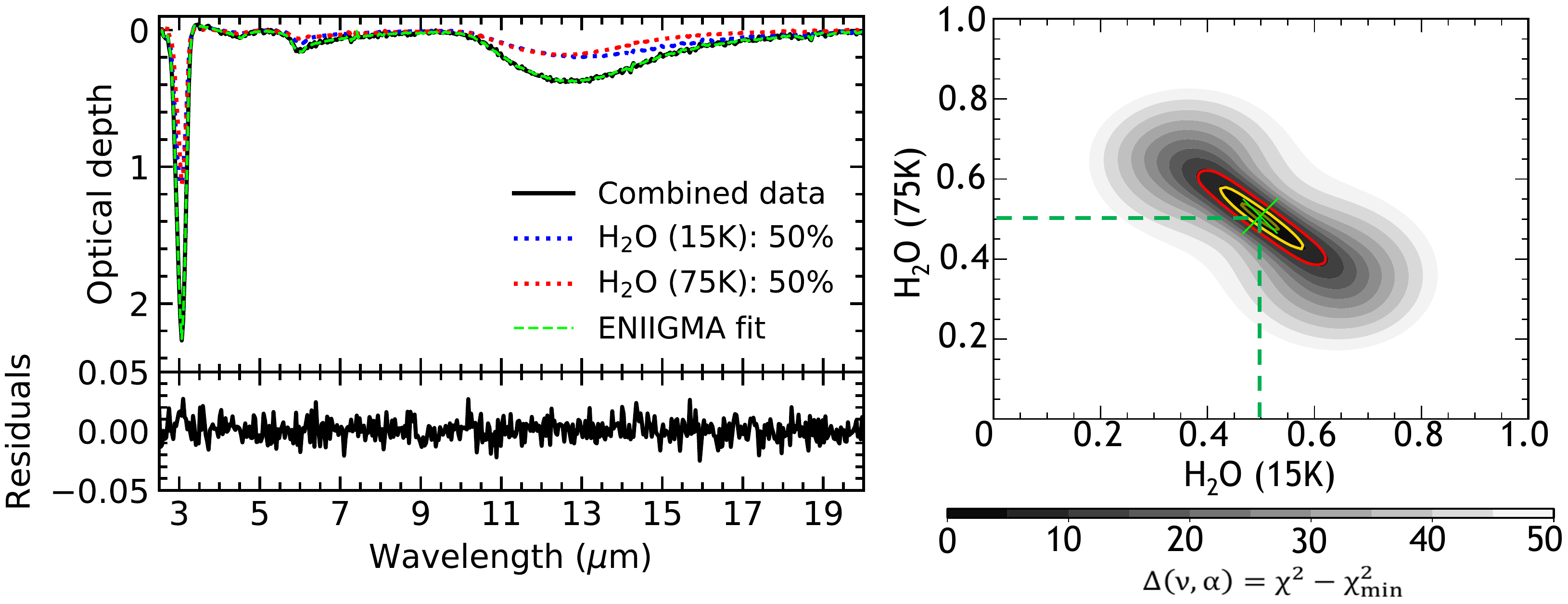}
    \includegraphics[width=\hsize]{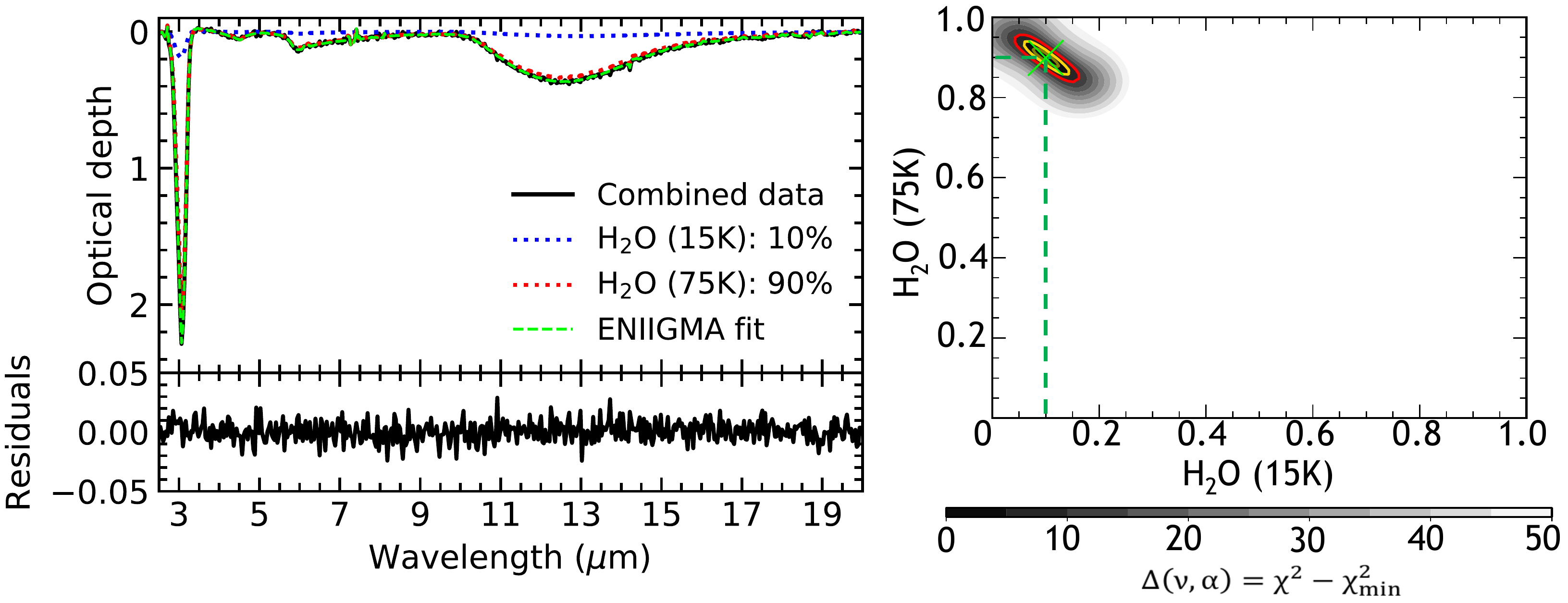}
      \caption{Same as Figure~\ref{Frac1}, but for H$_2$O at 15~K and 75~K.
              }
         \label{Frac2}
   \end{figure*}

A second check with water ice samples at (15~K) and (75~K) was also made. At high temperatures the water band shapes are significantly different than in the cases of 15~K and 40~K because of the transition from amorphous to cubic crystalline ice  \citep[][]{Hudgins1993}. In particular, the O$-$H band at 3~$\mu$m is gradually sharpened until the full crystalline ice is formed between 100$-$140~K \citep[][]{Sack1993}. As shown in the left panels of Figure~\ref{Frac2}, the \texttt{ENIIGMA} tool is able to identify the correct proportions of the two samples in the cases of 50:50 and 10:90, respectively. This is also indicated by the low fit residuals around 5\% in both situations.

The confidence intervals of the optimal solution are shown in the right panels of Figure~\ref{Frac2}. In these plots, the reduced area of the contours compared to Figure~\ref{Frac1} indicates that the solution is less degenerated. The reason is the differences of the band shapes of the cold and warm H$_2$O ice components. In the case of 50:50 proportion shown in the top right panel, the 3$\sigma$ confidence interval of both coefficients is around 20\%. On the other hand, in the bottom right panel, the degeneracy of the 10:90 proportion is 50\% and 10\%, respectively, which indicates that the H$_2$O (75~K) component is better constrained than the H$_2$O (15~K) component.

\begin{figure*}
   \centering
   \includegraphics[width=\hsize]{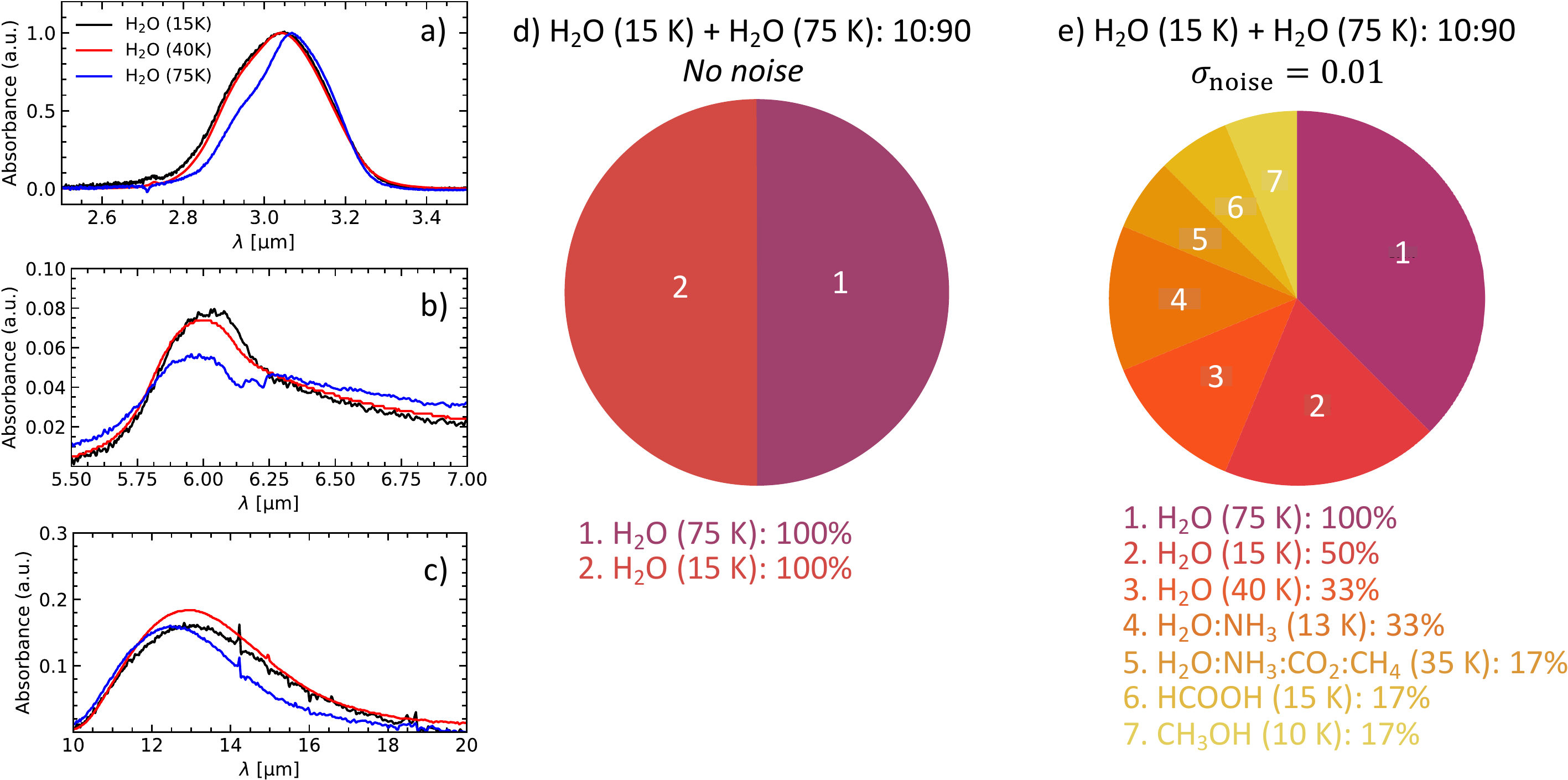}
      \caption{Degeneracy analysis for the combination of H$_2$O (15~K) + H$_2$O (75~K): 10:90. Panels {\it a.$-$c.} show respective stretching, bending and librational modes of H$_2$O ice at 15, 40 and 75~K. These panels highlight the temperature effect on the band shapes in the water ice IR spectra. Panel {\it d} shows the recurrence pie chart for the fit of the combined data H$_2$O(15~K):H$_2$O(75~K) with proportions 10:90, respectively and without adding Gaussian noise to the spectrum. Panel {\it e} shows the same as Panel {\it d}, but after adding Gaussian noise to the spectrum ($\sigma = 0.01$).}
         \label{Frac2_degen}
   \end{figure*}

After the global minimum solution is found, the degeneracy analysis of the ice composition is performed for the case with proportions of 10:90. Figures~\ref{Frac2_degen}a-c show the IR spectra of H$_2$O ice at 15, 40 and 75~K. As one can note, the spectral shape of the bands is similar between the samples at 15 and 40~K, and significantly different at 75~K because of the crystallization of the ice matrix via warm-up. An initial test has been done with the sum of the two H$_2$O ice samples at 15 and 75~K without Gaussian noise and the recurrence plot was created. The recurrence plot shown in panel {\it d} indicates that no degeneracy exists inside 3$\sigma$ confidence interval when the spectrum is absent of noise. On the other hand, if Gaussian noise is added to the summed spectrum, the degeneracy becomes evident. The dominant ice component, H$_2$O at 75~K is still present in all solutions inside the same confidence interval, whereas the recurrence of water component at 15~K is reduced to 50\% because of its lower contribution. The presence of other samples in this recurrence analysis indicate that the H$_2$O (15~K) can be replaced by one of those data, and still provide good solution inside 3$\sigma$ confidence interval.

\subsubsection{H$_2$O (15K) and CO:CH$_3$OH:CH$_3$CH$_2$OH (15K)}
\label{test3}
\begin{figure*}
   \centering
   \includegraphics[width=\hsize]{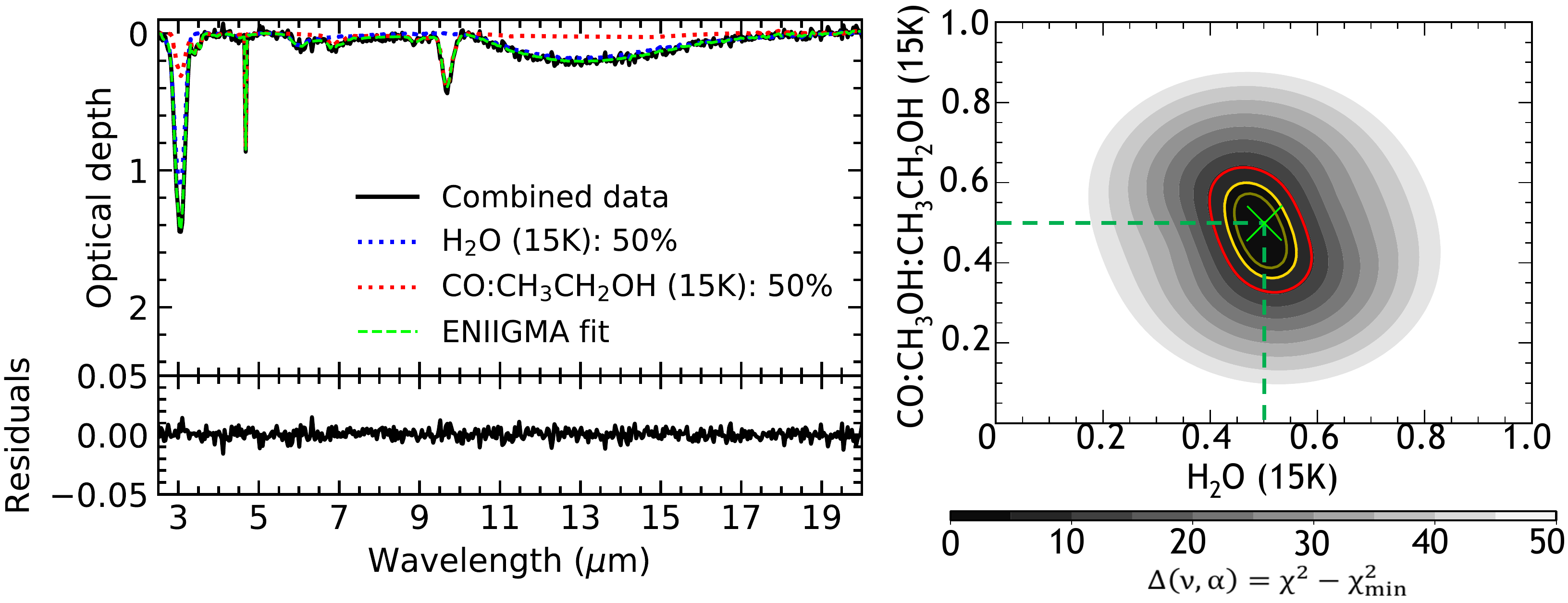}
        \includegraphics[width=\hsize]{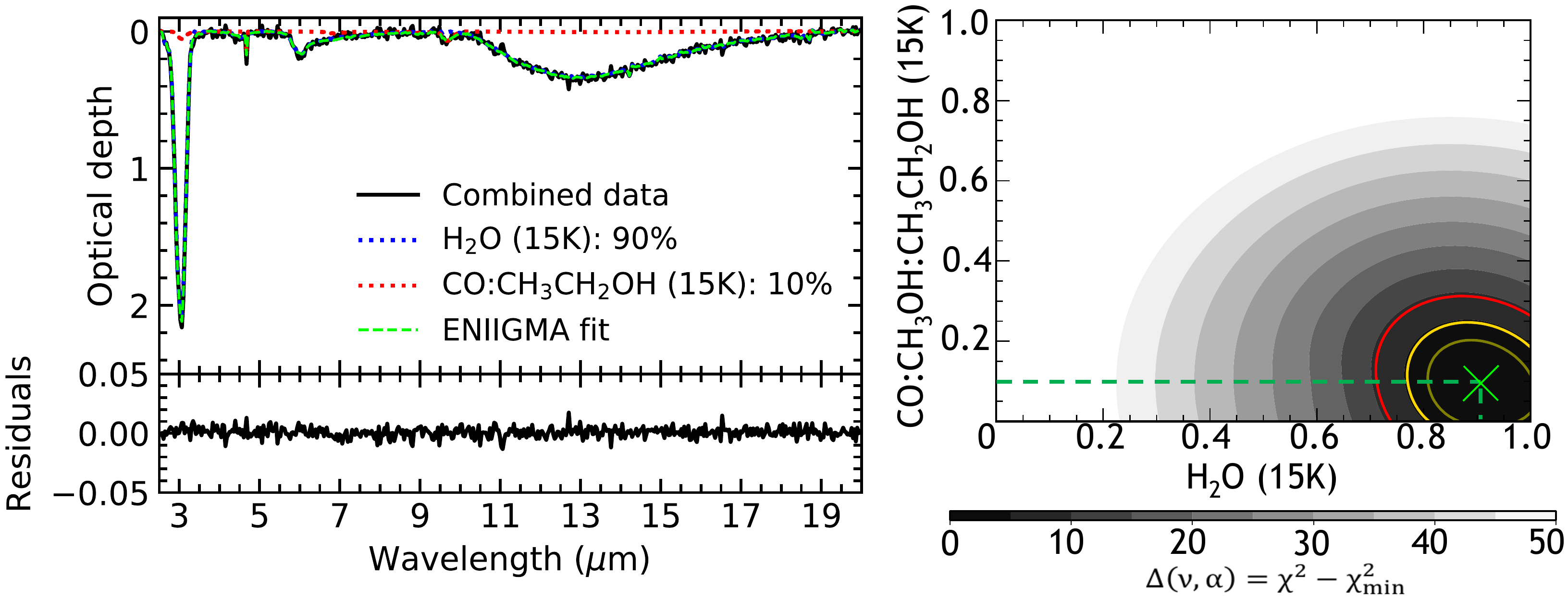}
      \caption{Same as Figure~\ref{Frac1}, but for H$_2$O and CO:CH$_3$CH$_2$OH at 15~K.
              }
         \label{Frac3}
   \end{figure*}

In this test, the combination of H$_2$O and the mixture CO:CH$_3$OH:CH$_3$CH$_2$OH (20:20:1) at 15~K is used. In general, these two ice data have different band peaks and shapes in the IR spectrum, although the O$-$H stretching modes of H$_2$O and CH$_3$OH overlaps around 3~$\mu$m. 
Another difference is that the samples do not have the same ice thickness. Figure~\ref{Frac3} shows the fits of the proportion 50:50. Because of the thickness difference, the water ice has a major contribution to the band at 3~$\mu$m compared to methanol, whereas at 4.67~$\mu$m and 8.86~$\mu$m the presence of the mixture sample becomes evident. The residuals of the fit are lower than 5\% and indicate a good match between fit and combined data. The confidence interval analysis of the coefficients, shown in the top right panel indicates that the water and mixture components vary in about 20\% and 40\%, respectively, inside 3$\sigma$.

The bottom left panel shows the fit of the proportions 90:10 for the water and mixture samples, respectively. In this case, the contribution of methanol at 3~$\mu$m is minimal, although it still visible around 9~$\mu$m. As seen from the residuals of the fit, the decomposition process was also able to identify the correct coefficients of the two samples. However, because of the small contribution of the mixture, the confidence interval analysis (bottom right panel) indicate that solutions considering only pure water are statistically significant as well. 

\subsubsection{Spurious features}
\label{test_spurious}
To test the effect of spurious bands that could be present in a real observational spectrum, we added emission and absorption features of non-ice compounds to the constructed mixture shown in Section~\ref{test3}. The emission features mimic the hydrogen lines at 3.04~$\mu$m (H\texttt{I} Pf$\epsilon$ 5$-$10), 3.30~$\mu$m (H\texttt{I} Pf$\delta$ 5$-$9), and 3.74~$\mu$m (H\texttt{I} Pf$\gamma$ 5$-$8). These lines are seen in spectra observed with ground-based telescopes when the standard star is not properly modelled \citep[e.g.,][]{Pontoppidan2004, Thi2006}. These emission profiles were created with the Python package \texttt{pysynphot}\footnote{\url{https://pysynphot.readthedocs.io}} \citep[][]{pysynphot2013} by adopting full width at half maximum of 50~$\AA$. For the absorption feature, we use the naphthalene (C$_{10}$H$_8$) spectrum taken from \citet{Hudgins1998}, that is the simplest polycyclic aromatic hydrocarbon containing two rings. Naphthalene has multiple bands between 3.20~$\mu$m and 3.44~$\mu$m, where features of methanol and ethanol are also present.

\begin{figure}
   \centering
   \includegraphics[width=\hsize]{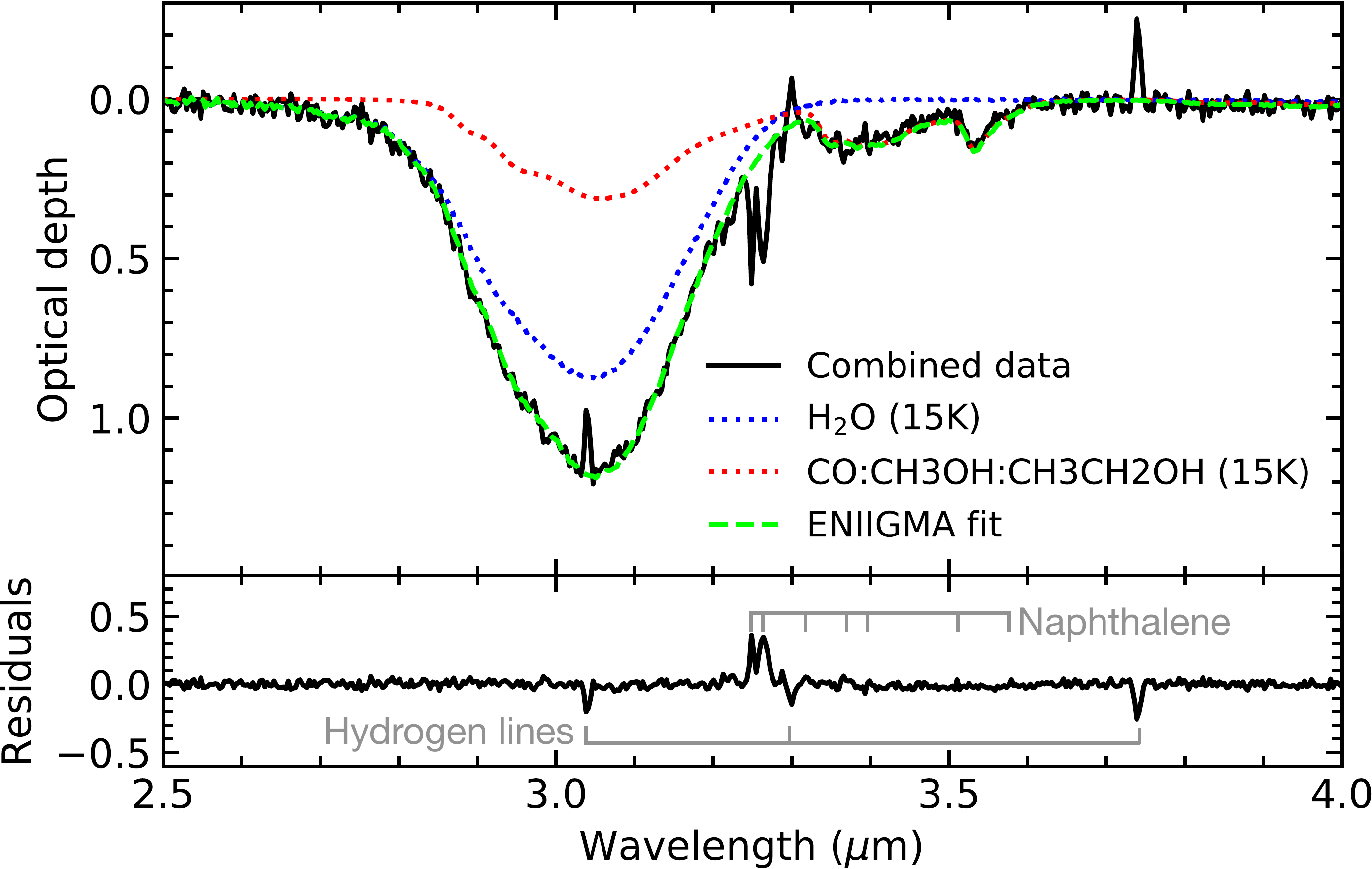}
      \caption{Spectral decomposition of the H$_2$O and CO:CH$_3$OH:CH$_3$CH$_2$OH constructed mixture containing spurious features mimicked with hydrogen lines and naphthalene bands.}
         \label{spurious_fetures}
   \end{figure}

Figure~\ref{spurious_fetures} displays the spectral decomposition of a constructed mixture containing ice bands (H$_2$O and CO:CH$_3$OH:CH$_3$CH$_2$OH) and spurious features. Once these emission and absorption features are located shortwards of 4~$\mu$m, the \texttt{ENIIGMA} fit is performed between 2.5~$\mu$m and 4.0~$\mu$m. This test shows that the \texttt{ENIIGMA} fitting tool identify correctly the ice bands and does fit the narrow features associated with spurious absorption and emission lines. In this example, the spectral profile between 3.2~$\mu$m and 3.5~$\mu$m is the most affected with spurious features, but those features do not change the shape of the ice bands. We highlight that in real observations, other spurious features might occur in the spectrum due to poorly continuum subtraction, as well as the effect of spectral glitches and baseline jumps. In these cases, a broad band fit of the spectrum covering other vibrational modes of the same molecule could minimize the effect of spurious features in the spectral decomposition with the \texttt{ENIIGMA} package.

\subsubsection{Synthetic ice spectrum}
\label{test4}
\begin{figure*}
   \centering
    \includegraphics[width=\hsize]{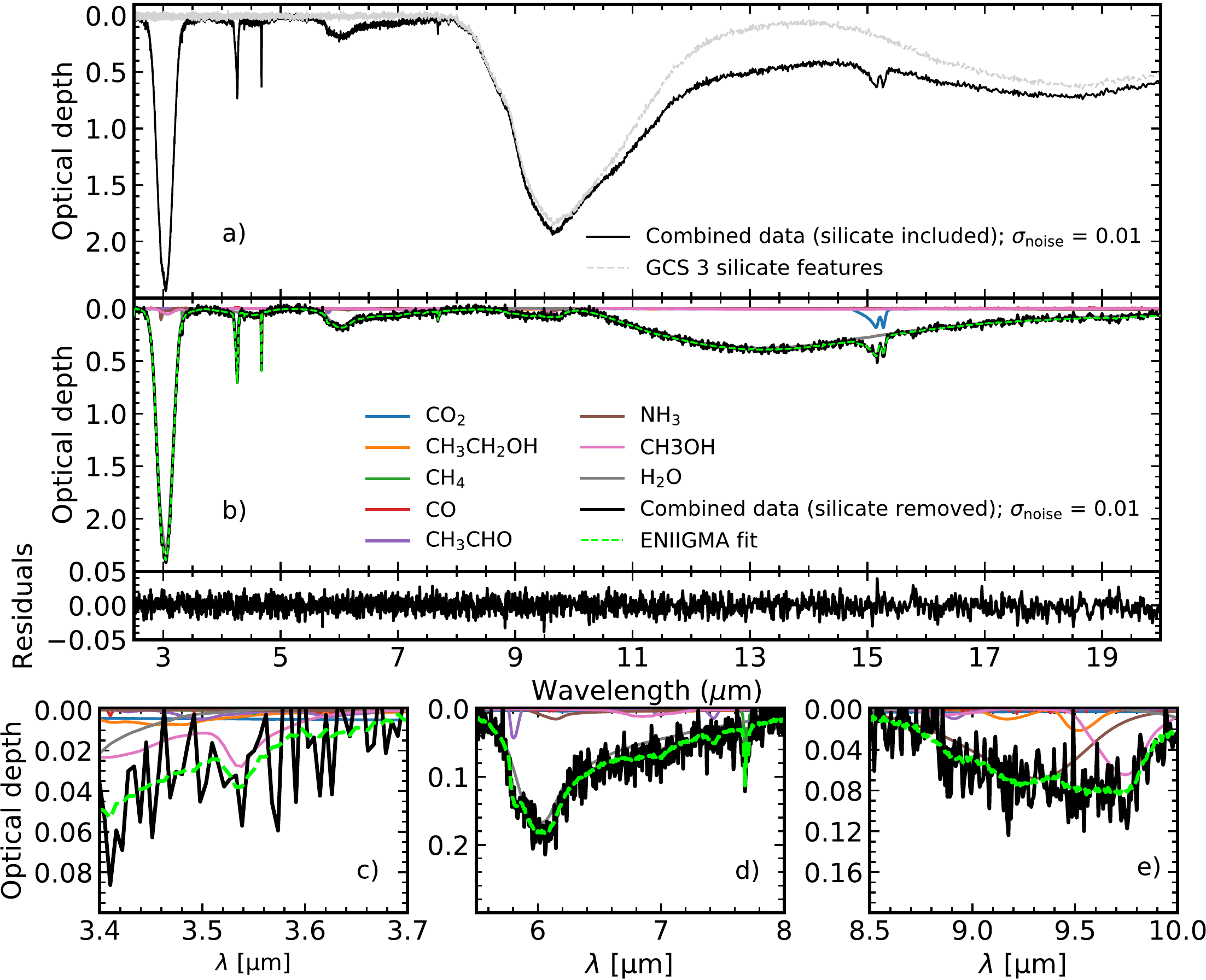}
      \caption{Spectral decomposition of the synthetic ice spectrum composed of 8 components of pure ice sample (black line) and Gaussian noise ($\sigma = 0.01$). {Panel a shows the synthetic spectrum along with the silicate band from 2.5$-$20~$\mu$m.} Panel b displays the synthetic spectrum without the silicate feature and residual plot.} The \texttt{ENIIGMA} fit is shown by the dashed green line and the ice components by the other colours. Panels c-e show small spectral ranges highlighting the contribution of shallow bands.
         \label{syntetic_ice_noise}
   \end{figure*}

Multiple ice features have been observed in the IR spectra of protostars \citep[e.g.,][]{Gibb2004, Zasowski2009}. For example, a linear combination of five empirical components, which are associated to more than one carrier, have been adopted to decompose the short spectral range between 5$-$8~$\mu$m \citep[][]{Boogert2008}. In this test, a synthetic ice spectrum is created based on the median ice composition derived by \citet{Oberg2011_spitzer} towards low-mass protostars, namely, H$_2$O:CO:CO$_2$:CH$_3$OH:NH$_3$:CH$_4$:XCN (1:0.29:0.29:0.03:0.05:0.05:0.003). However, the chemical specie XCN is replaced by 2\% of CH$_3$CHO and 1\% of CH$_3$CH$_2$OH with respect to the water ice to check the feasibility to identify these molecules with the \texttt{ENIIGMA} tool. The synthetic spectrum, shown in Figure~\ref{syntetic_ice_noise}a, is created with the linear combination of IR spectra of eight pure ices (temperatures: 10$-$15~K) listed in Table~\ref{ice_list_pure} and the silicate spectrum from GCS~3 taken from \citet{Kemper2004}. Also, a Gaussian noise of $\sigma$ equal to 0.01 is added to the spectrum. This highlights that a real spectrum might have such a strong silicate features, and that they must be subtracted off before the the spectral decomposition with the \texttt{ENIIGMA} tool. The procedure of silicate removal in an observed spectrum is shown in Sections~\ref{silc_removal} and \ref{EL29_removals}.

The initial guess adopted in this test is made of H$_2$O (40~K), NH$_3$ (10~K) and CH$_3$OH (75~K). Figures~\ref{syntetic_ice_noise}b$-$e show the result of the spectral decomposition. Although the expected global minimum solution was found by the three methods in Table~\ref{gen_op}, Method 1 provided less accurate coefficients in about 4$-$7\%. In the top panel, the fit is shown for the range between 2.5$-$20~$\mu$m. At this scale, only the major features are seen, such as H$_2$O (3~$\mu$m, 6~$\mu$m, 13~$\mu$m), CO$_2$ (4.27~$\mu$m, 15.53~$\mu$m), CO (4.67~$\mu$m) and CH$_4$ (7.673~$\mu$m). The residual plot shows that only the Gaussian noise is left after subtracting the data from the model, which indicates that both the samples and the coefficients were found by the \texttt{ENIIGMA} fitting tool. The bottom panels show a zoom-in of the regions where the weak features are detected. The left panel highlights the C$-$H stretching mode at 3.53~$\mu$m. The middle panel shows the contribution of CH$_3$CHO at 5.8~$\mu$m, and 7.4~$\mu$m, NH$_3$ at 6.2~$\mu$m and CH$_3$OH at 6.8~$\mu$m. The right panel shows the dominant absorption features of NH$_3$ at 9.35~$\mu$m and CH$_3$OH at 9.8~$\mu$m, with small contributions of CH$_3$CHO at 8.9~$\mu$m and the CH$_3$CH$_2$OH double peaks at 9.2~$\mu$m and 9.5~$\mu$m.

In summary, these tests show that the \texttt{ENIIGMA} can find the correct components and linear combination coefficients in the combined spectral data. However, Methods 2 and 3 provides better results than Method 1 when the number of components is large.

\section{Application to infrared spectra of the Class I protostar Elias 29}
To test the \texttt{ENIIGMA} fitting tool with a real source, an spectral analysis of the well-known Class I YSO, Elias~29 is carried out. This object is located in the Ophiuchus cloud, at an estimated distance of 135~pc \citep[][]{OrtizLeon2018}. Elias~29 is the brightest source in the $\rho$ Oph~E core and was observed in broad IR spectral range by the {\it Infrared Space Observatory}. Multiple absorption bands were securely and tentative detected in the IR spectrum of this protostar \citep{Gibb2004, Boogert2000, Rocha2015}. Millimeter observations toward Elias~29 show high abundances of simple molecules 
\citep[e.g., CO, H$_2$CO, HCO$^+$, SO, SO$_2$;][]{Boogert2002, Oya2019} and deficiency of complex molecules \citep[e.g., CH$_3$OH, HCOOCH$_3$, CH$_3$OCH$_3$;][]{Artur2019, Oya2019}. In particular, \citet{Oya2019} attribute the lack of organic molecules and richness in SO and SO$_2$ to the evolved nature of the source or the enhancement of the dust temperature above the CO freeze-out regime ($>$ 20~K) as suggested by \citet{Rocha2018} based on radiative transfer models. \citet{Artur2019} suggest that the non-detection of warm CH$_3$OH toward Elias~29 and other sources in Ophiuchus molecular cloud, is due to the absence of hot-core like region in the inner envelope close to the protostar or the due to higher temperatures in the precursor environments, as also discussed in the case of Corona Australis \citep[][]{Lindberg2014}.

\subsection{Continuum and optical depth calculation, and silicate feature extraction}
\label{EL29_removals}
\begin{figure}
   \centering
   \includegraphics[width=\hsize]{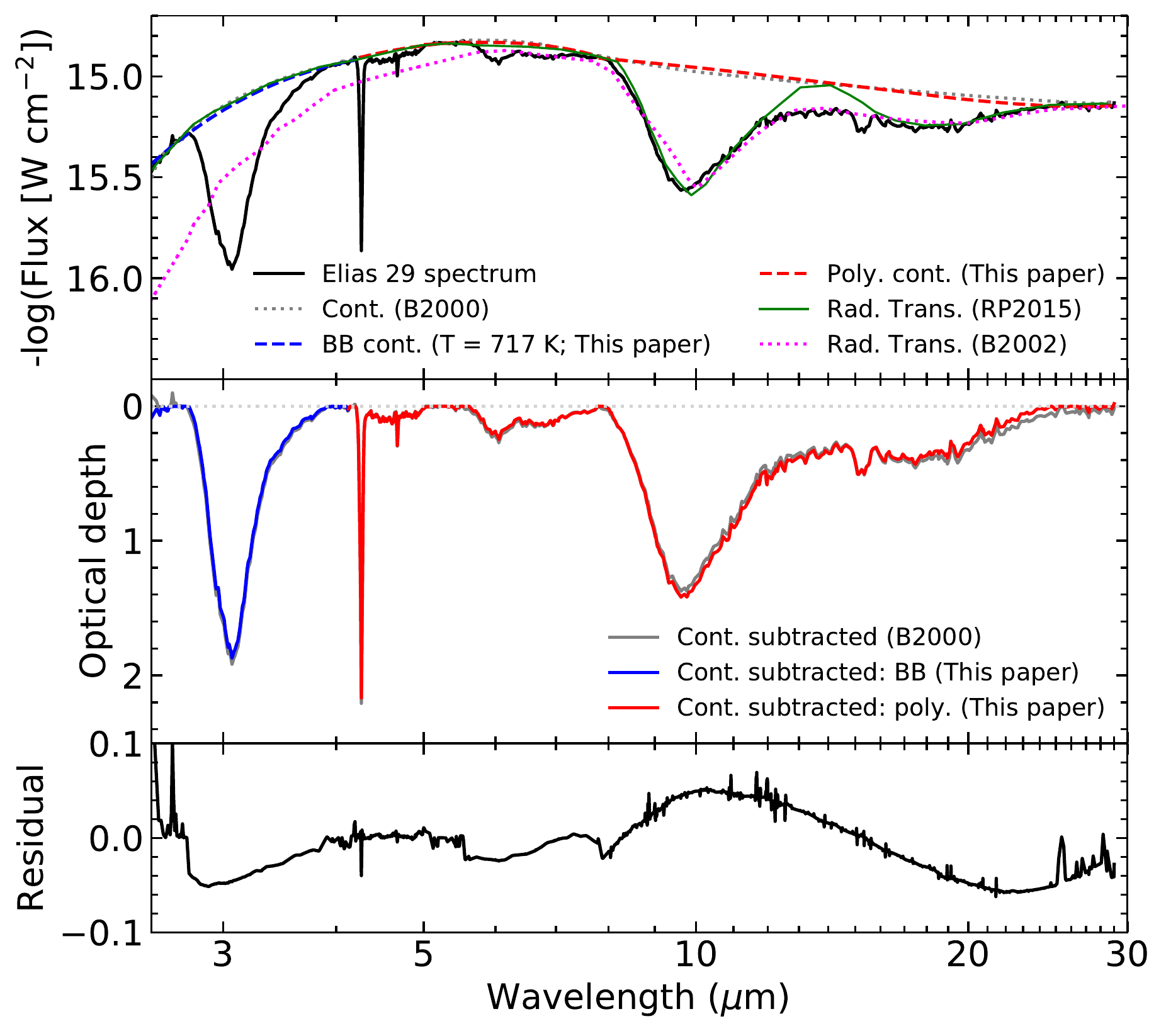}
      \caption{Continuum determination and optical depth calculation of the Elias~29 spectrum. {\it Top:} Blue and red dashed lines show, respectively, the blackbody and polynomial continuum calculation used in this paper. The grey dotted line shows the continuum determination from \citet{Boogert2000}, which is similar to the continuum in this paper. The dotted magenta line is the continuum calculation with radiative transfer modelling from \citet{Boogert2002}. The green solid line is the continuum calculation with radiative transfer modelling from \citet{Rocha2015}. In both cases, the silicate feature is red-shifted in the models compared with the observations.
      {\it Middle:} Grey line shows the optical depth scale of Elias~29 spectrum adopting the continuum from \citet{Boogert2000}, whereas blue and red lines show the same by respectively adopting blackbody and polynomial continuum derived in this paper.
      {\it Bottom:} Difference in optical depth estimates from this paper and \citet{Boogert2000}.}
         \label{contElias29}
   \end{figure}

In this paper, the continuum SED of Elias~29 has been determined with a blackbody function shortwards of 4.2~$\mu$m and a low-order polynomial function between 4.1~$\mu$m and 30~$\mu$m. In the former case, the spectrum between the ranges 2.5$-$2.8~$\mu$m and 3.9$-$4.2~$\mu$m are absent of ice features and thus were used to constrain the fitting. In this case, a temperature of 717~K provided the best match with the observations. Longwards of 4.1~$\mu$m, a third-order polynomial fitting passing through spectral ranges without ice features (4.1$-$4.25~$\mu$m, 4.3$-$4.5~$\mu$m, 5.3$-$5.5~$\mu$m and 25$-$30~$\mu$m) has been adopted. \citet{Pontoppidan2005} mention that although this method has some uncertainty involved, the shape of the bands are not affected even though the total optical depth is uncertain by up to 20\%. The top panel of Figure~\ref{contElias29} shows the result of this fit and other calculations in the literature. The fits from \citet{Boogert2000} and \citet{Rocha2015} agrees better with the continuum SED calculated in this paper, whereas the fit from \citet{Boogert2002} mismatch the observations.

Once the continuum SED is determined, the observational optical depth is calculated with Equation~\ref{tau_obs_eq}, and the result is shown in the middle panel of Figure~\ref{contElias29}. The continuum methods used in this paper and \citet{Boogert2000} results in similar optical depths. The differences between the two methods are shown in the bottom panel of Figure~\ref{contElias29} and are less than 10\% in the range between 2.5$-$30~$\mu$m. A similar variation is reported by \citet{van_Breemen2011} for the band at 9.8~$\mu$m, when different continuum models are assumed for the objects StRs~164 and SSTc2d\_J163346.2$-$242753. Most importantly, they concluded that the shape of the 9.8~$\mu$m bands remains virtually unaffected, which is in agreement with the result shown in this work. On the other hand, the effect of alterations in the shape of the silicate band at 18~$\mu$m are much less constrained.

\begin{figure}
   \centering
   \includegraphics[width=\hsize]{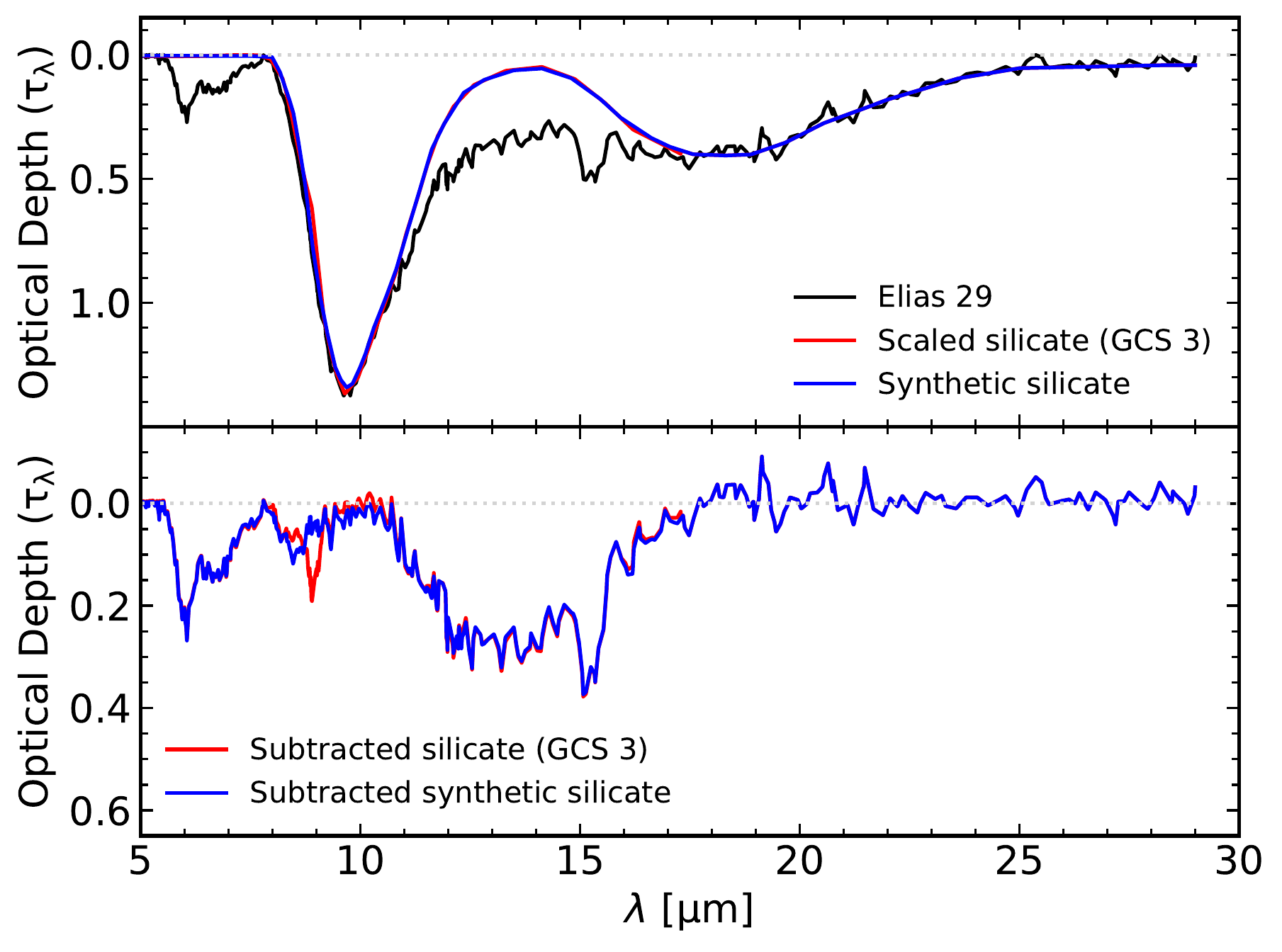}
      \caption{Silicate extraction procedure in the Elias 29 spectrum. {\it Top:} Black line shows the optical depth scale of the Elias 29 spectrum calculated in this paper. Red and blue lines show, respectively, the scaled CGS~3 and synthetic silicate bands at 9.7~$\mu$m and 18~$\mu$m. {\it Bottom:} Red and blues lines show the Elias 29 mid-IR optical depth after silicate band extractions with scaled CGS~3 and synthetic silicate profiles, respectively.}
         \label{OD_silicate_comp}
   \end{figure}

The silicate absorption in the spectrum of Elias~29 is removed following the procedure described in Section~\ref{silc_removal}. It is observed in the top panel of Figure~\ref{OD_silicate_comp} that the GCS~3 silicate feature resembles well the absorption band seen toward Elias~29, suggesting they could have similar mineralogy. In this particular case, the modification of the 9.7~$\mu$m band shape is minimal as can be noted in the result of the silicate extraction in the bottom panel of Figure~\ref{OD_silicate_comp}. The major difference in the two methods is only the absorption excess at 8.8~$\mu$m, previously discussed in Section~\ref{silc_removal}. With the synthetic silicate band, this leftover residual is subtracted off. Figure~\ref{OD_silicate_comp}a also shows an absorption excess between 11$-$18~$\mu$m between the broad silicate bands at 9.8~$\mu$m and 18~$\mu$m that is clearly seen in the silicate feature subtracted spectrum (Figure~\ref{OD_silicate_comp}b.) Because of the uncertainties in the continuum and in the silicate band profile at 18~$\mu$m, some effect could be added to this band with silicate subtraction. However, \citet{Boogert2008} and \citet[][]{Zasowski2009} convincingly shows that this entire feature is associated with the H$_2$O ice libration mode.

\subsection{Spectral decomposition}
\label{Spec_decomp}
\begin{figure*}
   \centering
   \includegraphics[width=\hsize]{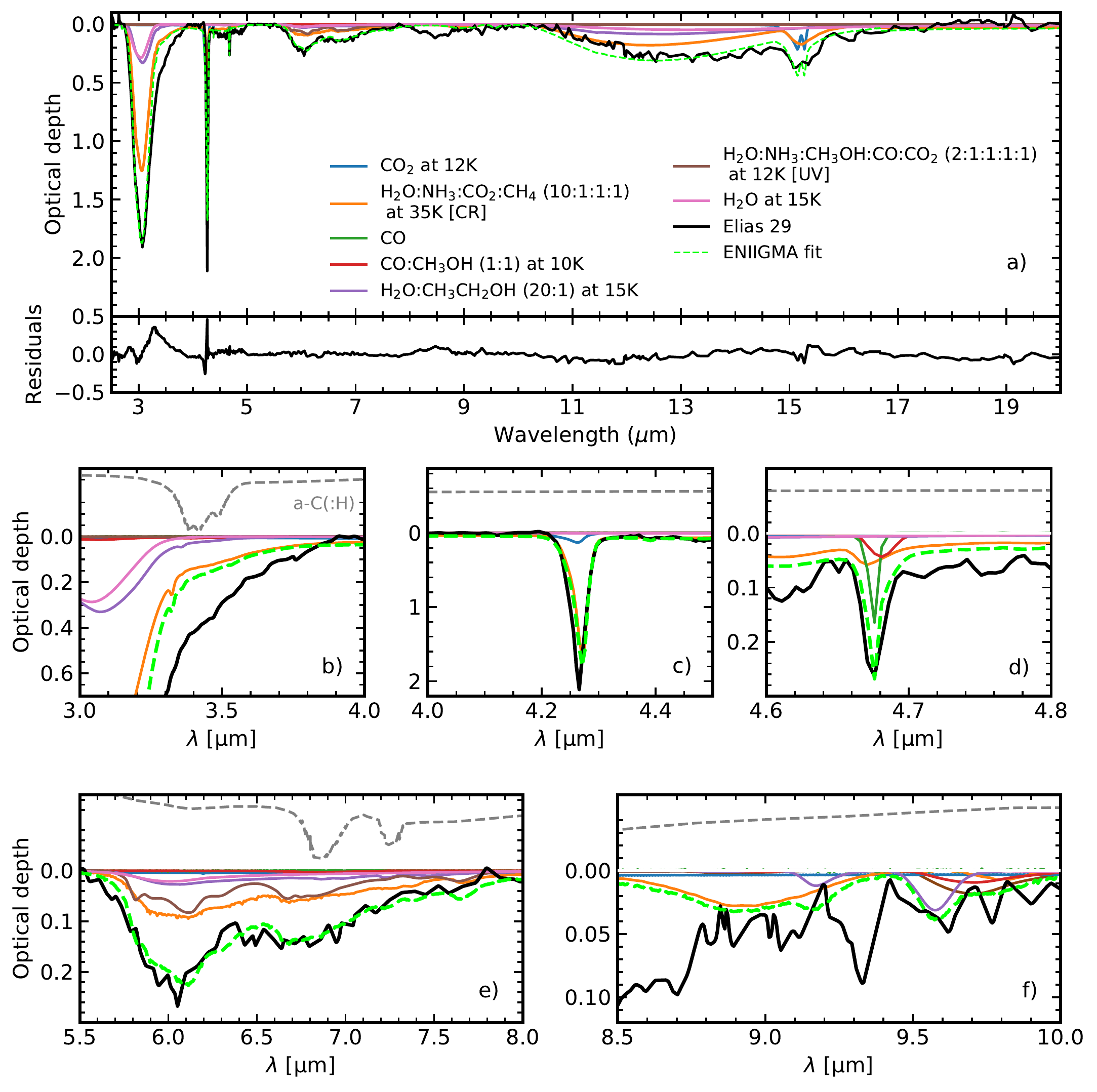}
      \caption{Spectral decomposition of the mid-IR spectrum of Elias~29 protostar. {\it a):} Optical depth scale of the observed spectrum between 2.5$-$20~$\mu$m, and the seven ice components used to fit the observed spectrum. The best fit is shown by the dashed green line. The residual spectrum is shown below this panel. {\it b-f:} Zoom-in of short spectral ranges highlighting the contribution of small features. In these plots, the spectral features of amorphous carbon, a-C(:H), that were not used in the spectral decomposition, are shown by the dashed grey lines above the Elias~29 spectrum and the ice data components. This illustrates potential contribution of features other than vibrational modes of ice species.} Panel {\it b} shows the L-band red wing absorption excess that is not entirely reproduced by the ice components (see Section~\ref{Spec_decomp}). Panels {\it c} and {\it d} display the CO$_2$ and CO features. Panel {\it e} shows important contributions of UV and cosmic-ray processed samples. Finally, Panel {\it f} highlights the contribution of the broad NH$_3$ ice feature at 9~$\mu$m, CH$_3$CH$_2$OH at 9.6~$\mu$m and CH$_3$OH at 9.8~$\mu$m.
         \label{EL29_decomp}
   \end{figure*}

The Methods~1$-$3 in Table~\ref{gen_op} were used to decompose the IR spectrum of Elias~29. The initial guess adopted in this study is composed of the pure ice samples H$_2$O, CO, CO$_2$ and NH$_3$. After selecting the best candidates for the solution, the entire sample is combined in groups of 7 components. The best global minimum was found in Method 3, and the result is shown in Figure~\ref{EL29_decomp}a$-$f. In Panel a, the spectral decomposition in the range between 2.5$-$20~$\mu$m highlights the fitting of the most prominent bands, whereas the Panels b$-$f bottom panels show zoom-in of small spectral ranges and the contributions of shallow features. The spectral features of a$-$C(:H), taken from \citet{Alata2014}, are also shown in these panels to illustrate potential contributions of carbonaceous grains as discussed by \citet[][]{Jones2012, Jones2016}.

The global minimum solution in the Elias~29 spectral analysis is given by the combination of pure and mixed ice samples. The dominant component in this broadband analysis is H$_2$O:NH$_3$:CO$_2$:CH$_4$ (10:1:1:1) at 35~K and processed by heavy ions, simulating the effects of cosmic rays. This sample has significant contributions at 3~$\mu$m, 6~$\mu$m and 11$-$18~$\mu$m and 15~$\mu$m. The second prominent component is CO$_2$ with vibrational modes at 4.27~$\mu$m and 15~$\mu$m. Pure and mixed H$_2$O components also contribute about 20\% of the remaining absorption at 3~$\mu$m. The shallow features observed in this spectral decomposition are CO, NH$_3$, CH$_4$, CH$_3$CH$_2$OH.

An old-standing problem in the analysis of mid-IR spectra of YSOs and background sources is that the band at 6~$\mu$m is deeper than what is expected from the 3~$\mu$m and 13~$\mu$m H$_2$O ice bands \citep[e.g.,][]{Cox1989, Gibb2000, Boogert2008, Bottinelli2010, Boogert2011}. A potential explanation for the absorption excess at 6~$\mu$m is the presence of organic refractory material \citep{Gibb2002}, although it lacks independent spectroscopic confirmation \citep[][]{Boogert2015}. Another open question in this context is the correct carrier of the band at 6.85~$\mu$m. Although the most likely candidate is NH$_4^+$ \citep[][]{Schutte2003}, it lacks a convincing profile fit as pointed-out by \citet{Boogert2015}.

In this paper, the global optimization with the \texttt{ENIIGMA} tool provided a good fit of the absorption bands observed in the Elias~29 spectrum between 2.5$-$20~$\mu$m. The feature at 3~$\mu$m associated with the O$-$H vibrational mode is fitted by multiple ice sample containing water. However, there is an absorption excess in the Elias~29 spectrum that is not fitted with the ice samples in this case (Figure~\ref{EL29_decomp}b). One of the explanations for this excess is the absorption caused by ammonia hydrates \citep[e.g.,][]{Knacke1982, Dartois2001}. Another cause is attributed to the light scattering due to large grains \citep[][]{Smith1989}. \citet{Jones2016} also suggest that compounds containing the carbonyl functional group (C$=$O) present in the dust before or during the formation of the ice mantle, could be a potential explanation for this absorption excess. Figure~\ref{EL29_decomp}b shows that a$-$C(:H) could contribute to the large absorption excess around 3.5~$\mu$m, although the absorption features at 6.85~$\mu$m and 7.24~$\mu$m lacks visual matching with the observations (Figure~\ref{EL29_decomp}e). Because of the low spectral resolution, this does not disregard the presence of a small fraction of a$-$C(:H) in the dust grains toward Elias~29. However, \citet[][]{Boogert2000} argue that for the case of Elias~29, the scattered light due to large grains is the likely explanation because of the good matching of this red wing obtained with large icy grains.

The CO$_2$ stretching component at 4.27~$\mu$m (Figure~\ref{EL29_decomp}c) is found to be a combination of pure and mixed CO$_2$. However, the peak optical depth in the fit does not match with the observations because it is also constrained by the CO$_2$ bending mode at 15.1~$\mu$m. Adding more CO$_2$ to the feature at 4.27~$\mu$m would lead to absorption excess at 15.1~$\mu$m. On the other hand, the CO ice band at 4.67~$\mu$m (Figure~\ref{EL29_decomp}d) is reproduced in this study by three components, namely, pure CO ice, CO:CH$_3$OH and CO formed with the cosmic-ray processing of H$_2$O:NH$_3$:CO$_2$:CH$_4$ (10:1:1:1) at 35~K, which is in agreement with the literature. Previous studies of the structure of CO ice band \citep[e.g.,][]{Pontoppidan2003, Thi2006, Perotti2020}, have shown that this band is well fitted by three Gaussian components or a combination of laboratory data and analytical functions. Such components are associated with pure CO, and CO mixed in polar ice matrices \citep[e.g., H$_2$O and CH$_3$OH;][]{Cuppen2011}.

The broadband between 5.5$-$8.0~$\mu$m (Figure~\ref{EL29_decomp}e) is well fitted by combining multiple components. In particular, two ice samples 
processed by UV \citep[H$_2$O:NH$_3$:CH$_3$OH:CO:CO$_2$;][]{Caro2003} and cosmic rays \citep[H$_2$O:NH$_3$:CO$_2$:CH$_4$][]{Pilling2010ammonium} significantly contribute to fit the Elias~29 spectrum. The ammonium ion (NH$_4^+$) is a common byproduct of the ice processing in these two cases, and is one of the proposed carriers of the band at 6.85~$\mu$m. The significant contribution of these two IR ice spectrum might indicate that the chemical evolution of Elias~29 is mostly induced by energetic processing of the ice mantles.

In the presented spectral analysis, CH$_3$CH$_2$OH mixed in H$_2$O ice is a likely carrier of absorption bands at 7.5~$\mu$m and 9.6~$\mu$m (Figure~\ref{EL29_decomp}e,f). Although the ethanol bands cannot be resolved at 7.5~$\mu$m with {\it ISO} observations, the band at 9.6~$\mu$m could indicate a tentative detection of this complex organic molecule toward Elias~29. This result also shows the contribution of methanol ice at 9.75~$\mu$m. Most importantly, this result shows that the band between 9.5$-$10~$\mu$m might not be only associated with methanol as suggested previously \citep[e.g.,][]{Boogert2008, Bottinelli2010}, but rather that it is due to the presence of other alcohols in ices as discussed by \citet{Jones2016}. Consequently, the reduced amount of methanol in ices would contribute to partially solve the conundrum of the low gas-to-ice ratio observed toward star-forming regions \citep[$\sim$ 10$^{-4}$;][]{Oberg2009a, Perotti2020, Perotti2021}, as also conjectured by \citet[][]{Jones2016}. However, we stress that high resolution spectral data are required to unambiguously detect complex molecules in ices beyond methanol. 

\subsection{Recurrence analysis of the spectral decomposition}
\begin{figure}
   \centering
   \includegraphics[width=8cm]{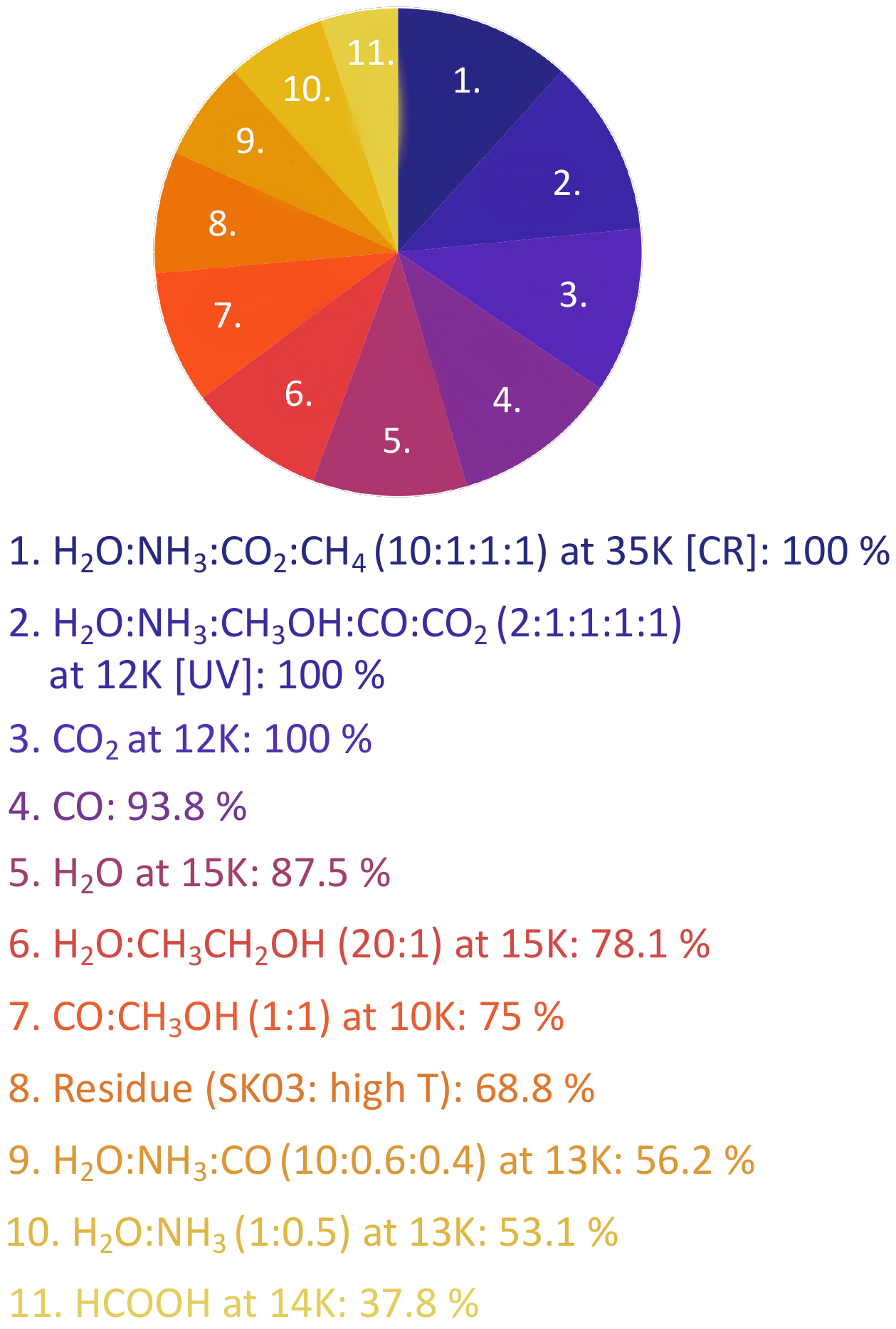}
      \caption{Recurrence pie chart of the IR ice data contributing to solutions inside 3$\sigma$ confidence interval ($\Delta \chi^2 \leq 18.5$). The size of each piece is proportional to the contribution of each sample. }
         \label{Rec_EL29}
   \end{figure}

Because the ice spectral features vary with the physical and chemical environment such as composition, temperature, and radiation field, the fits of observational IR spectra of YSOs are often degenerate. The recurrence pie chart (Section~\ref{Stats}) is used to verify uniqueness of the solution found with the \texttt{ENIIGMA} and the result is displayed in Figure~\ref{Rec_EL29}.

This analysis shows that the samples 1$-$7 have recurrence above 75\%, and are also the components providing the global minimum solution as seen in Figure~\ref{EL29_decomp}. The samples 1$-$3 have a recurrence of 100\%, and thus are solutions that cannot be replaced. Additionally, they have a significant contribution to the strong absorption features in the Elias~29 spectrum. The CO sample has a recurrence of 93.8\% since the C-O stretching mode at 4.67~$\mu$m can be combined with other solutions containing CO, such as sample 1 (CO formed after processing; see Section~\ref{Spec_decomp}) and samples 7 and 9. The recurrence of pure H$_2$O (87.5\%) indicates that this component is necessary for the global minimum solution, but not crucial. Other samples containing H$_2$O (e.g., 6, 9 and 10), can also be adopted to provide a good fit of the water absorption bands. Sample 6, is the only component containing CH$_3$CH$_2$OH. However, because this sample is dominated by water ice, that can be replaced by another component, its recurrence is 78.1\%. Sample 7 has a recurrence of 75\% and is important to fit the red wing of the absorption band at 4.67~$\mu$m. 

The samples 8$-$11 are less recurrent (37.7$-$68.8\%), although still inside the 3$\sigma$ confidence interval. In particular, sample 8 contains a strong NH$_4^+$ feature that could replace component 2 in the fits. However, it is not part of the global minimum solution because the bandwidth at 6.85~$\mu$m of component 8 is lower than in component 2, which provided a good match with the observational spectrum. In the cases of samples 9 and 10, they provide less accurate fit compared to components 1, 5, and 6. The less recurrent component, HCOOH, has been tentatively detected towards Elias~29. Despite formic acid is not part of the global minimum solution, it is not excluded as a potential carrier to fit the Elias~29 spectrum.

\subsection{Ice column density}
\begin{figure}
   \centering
   \includegraphics[width=\hsize]{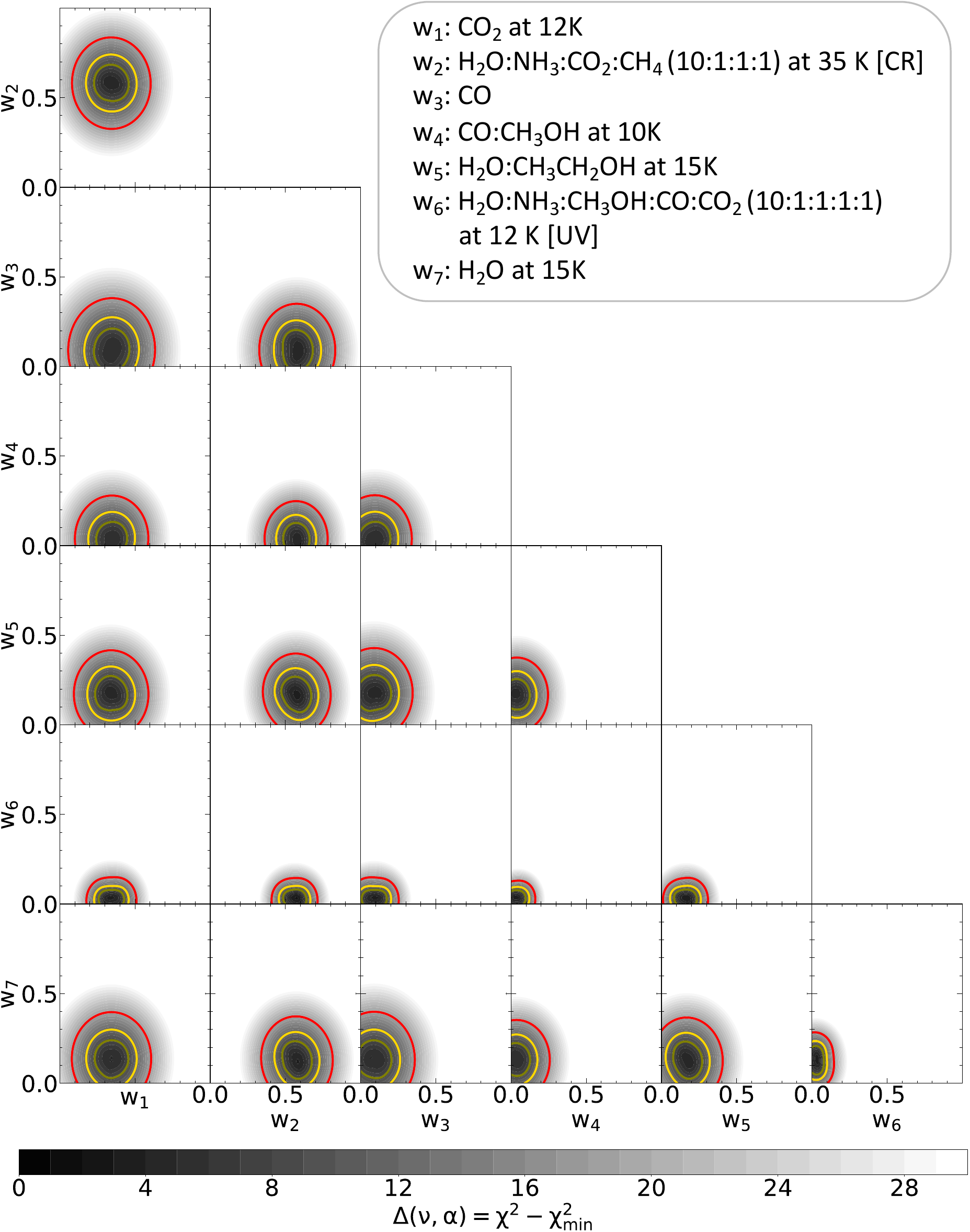}
      \caption{2D $\Delta \chi^2$ maps in grey-scale showing the correlation between seven ice laboratory data used to decompose the Elias~29 spectrum. The green, yellow and red contours indicate 1$-$3$\sigma$ confidence intervals, respectively. For seven components, they are defined for $\Delta \chi^2$ equal to 8.38, 12.02 and 18.48.}
         \label{CI_EL29}
   \end{figure}

Ice bands can be blended in the IR spectrum, and thus the selection of clear absorption features is essential to provide accurate quantification of the ice column densities. With the spectral decomposition used in this paper, the bands associated with different molecules were isolated, and their ice column densities calculated using Equation~\ref{CD_eq} and the band strengths listed in Table~\ref{ice_bs}. The derived values are shown in Table~\ref{ice_cd_el29} and are compared with previous calculations by \citet{Boogert2000}. The lower and upper bounds are calculated from the 3$\sigma$ confidence interval analysis shown in Figure~\ref{CI_EL29}. Table~\ref{ice_cd_el29} also shows the median ice column densities calculated from the histograms shown in Appendix~\ref{Ap_hist}. These values take into account all solutions found by the \texttt{ENIIGMA} tool inside 3$\sigma$ confidence interval.

\begin{table*}[h]
\caption{\label{ice_cd_el29} Ice column densities towards Elias 29. Tentative detection in the literature are shown as 3$\sigma$ upper limit and non-detection as ``NA''.}
\renewcommand{\arraystretch}{1.5}
\centering 
\begin{tabular}{lcccc}
\hline\hline
Specie &  $N_{\rm{ice}}$ [$10^{17}$ cm$^{-2}$] & \multicolumn{2}{c}{This paper: $N_{\rm{ice}}$ [$10^{17}$ cm$^{-2}$]}\\
\cline{3-4}
       & \citet{Boogert2000} & Global minimum &  Median (Histogram$^b$)\\
\hline
H$_2$O &  34 (6)   & 33.11$_{21.09}^{41.13}$ & 35.42$_{25.23}^{41.87}$\\
NH$_3$ &  < 3.5   & 1.01$_{0.75}^{1.47}$ & 2.42$_{0.43}^{3.88}$\\
CO &  1.7 (0.3)   & 1.55$_{0.73}^{1.81}$ & 1.42$_{1.23}^{1.87}$\\
CO$_2$ &  6.7 (0.5)   & 5.22$_{3.4}^{7.13}$ & 4.42$_{2.23}^{6.87}$\\
CH$_4$ &  < 0.5   & 0.38$_{0.17}^{0.51}$ & 0.42$_{0.02}^{1.12}$\\
H$_2$CO &  < 0.6   & 0.38$_{0.13}^{0.42}$ & 0.27$_{0.03}^{0.52}$\\
CH$_3$OH &  < 1.5   & 0.86$_{0.06}^{1.16}$ & 1.70$_{0.21}^{2.38}$\\
CH$_3$CH$_2$OH &  NA   & 0.08$_{0.03}^{0.11}$ & 0.12$_{0.02}^{0.40}$\\
NH$_4^+$ &  1.01 (0.3)$^a$  & 1.15$_{0.3}^{0.72}$ & 1.02$_{0.23}^{3.12}$\\
HCOOH &  < 0.3   & NA & 0.55$_{0.13}^{0.91}$\\
OCS &  < 0.015   & NA & NA\\
OCN$^-$ &  < 0.067   & NA & NA\\
\hline
\end{tabular}
\tablefoot{$^a$ \citet{Boogert2008}. $^b$ From the histograms shown in Appendix~\ref{Ap_hist}.}
\end{table*}

Water ice has multiple bands detected in the Elias~29 spectrum. In this paper, the water libration mode at 14~$\mu$m is used in the calculation as it easily separated from the CO$_2$ vibrational mode at 15~$\mu$m. The bands at 3~$\mu$m and 6~$\mu$m are blended with strong features of some molecules such as CH$_3$OH, HCOOH and processed ice, which makes difficult the correct extraction of the water band. In the cases of CO$_2$ and CO, the respective bands at 4.27~$\mu$m and 4.67~$\mu$m were used to derive the column densities. However, the pure and mixed CO$_2$ components found in the decomposition were not sufficient to reproduce the observed absorption band, and the $N_{\rm{CO_2}}$ ice column density in this paper is 28\% lower than found by \citet[][]{Boogert2000}. On the other hand, the $N_{\rm{CO}}$ values match well within the uncertainties. The ice column densities of calculated in this paper for H$_2$CO (5.83~$\mu$m), CH$_3$OH (6.83~$\mu$m), CH$_4$ (7.67~$\mu$m) and NH$_4^+$ (6.85~$\mu$m) are also in agreement with \citet{Boogert2000}.) 

From the \texttt{ENIIGMA} fits, CH$_3$CH$_2$OH contributes to the absorption band at 9.6~$\mu$m, and has abundances with respect to H$_2$O and CH$_3$OH ices of, respectively, 0.25\% and 10\%. Other molecules such as HCOOH, OCS and OCN$^-$ were not found in this work.

\section{Limitations}
\label{sec_limitations}
The methodology adopted in the \texttt{ENIIGMA} fitting tool seems to work well to decompose ice bands in the spectra of YSOs. However, we mention in this section the limitations of this tool, thereby listing opportunities for further development.

\underline{First:} poorly subtracted continuum or silicate band would add spurious features to the spectrum, that could bias the \texttt{ENIIGMA} spectral decomposition. For example, due to inherent uncertainties in the broad 18~$\mu$m amorphous silicate band subtraction, the shape of the H$_2$O libration and CO$_2$ bands around 13~$\mu$m could be affected. We thus recommend fitting other water and carbon dioxide absorption bands when possible to increase the reliability of the spectral decomposition method. We highlight that the approaches included in the \texttt{ENIIGMA} fitting tool are simplified techniques that have been successful while dealing with IR observations containing ice bands \citep[e.g.,][]{Boogert2008, Bottinelli2010}. Nevertheless, more accurate approaches apart from the \texttt{ENIIGMA} package could be adopted, instead. For example, a separate subtraction of silicate and other dust features could be performed with other computational tools \citep[e.g., THEMIS code;][]{Jones2017}, and the silicate subtracted spectrum decomposed with the \texttt{ENIIGMA} fitting tool.

\underline{Second:} the solutions found with the \texttt{ENIIGMA} fitting tool rely on the amount and diversity of data stored in the \texttt{ENIIGMA} database, which is flexible to be enlarged or reduced. While adding data to the database, we recommend caution to add properly baselined laboratory data, once inaccurate baselines would bias the results as well.

\underline{Third:} experimental works show that both temperature and chemical environment change the shape and peak position of ice bands \citep[e.g.,][]{Schutte2003, Dawes2016}. This imposes an intrinsic caveat to the methodology adopted in the \texttt{ENIIGMA} fitting tool, i.e., the effects of intermolecular interactions within the ice are limited to the mixtures present in the \texttt{ENIIGMA} database. It is worth noting, however, that to the best of our knowledge, there is no analytical method that systematically explains such a change in the ice bands with the chemical environment. Works by \citet{Bonfim_Pilling2018} started addressing this issue using quantum chemistry calculations, where an average dielectric constant simulates a generic mixture. One way to overcome this limitation is to add reliable experimental data of icy mixtures to the database.

\underline{Fourth:} the current version of the \texttt{ENIIGMA} fitting tool does not account for spectral band shape changes caused by the geometry and properties of the grain where the ice is formed in interstellar conditions.

\section{Summary}
In this paper, the \texttt{ENIIGMA} fitting tool is introduced as a new approach to perform spectral decomposition of ice features in the IR using genetic modelling algorithm. Python functions to perform continuum calculation and silicate removal in YSOs spectra before the decomposition are also available in the tool. Additionally, a post-processing module dedicated to statistical analysis of the solutions is added to the capabilities of the \texttt{ENIIGMA} tool. A compilation of 103 laboratory ice spectrum split into pure and mixed, as well as thermally or energetic processed ices compose an internal database used in the spectral analysis. 

In order to test all capabilities of the \texttt{ENIIGMA} fitting tool, fully blind and fully sighted tests were performed on known species. The statistical module was applied to derive the confidence intervals of coefficients in the global minimum solution. Moreover, the code was used to successfully decompose a synthetic ice spectrum containing different fractions of chemical species with respect to the water, as well as constructed mixture containing spurious absorption and emission features.

A final test on a real source showed that the \texttt{ENIIGMA} successfully decomposed the Elias~29 spectrum with 7 components, which correspond to pure and mixed ice samples. From this analysis, multiple molecules were identified, including a tentative detection of CH$_3$CH$_2$OH at 9.6~$\mu$m. Moreover, the ice column densities and their lower and upper limits derived with the \texttt{ENIIGMA} tool from the global minimum solution and histogram analysis are in agreement with the previous estimations in the literature.

The tests performed with synthetic and observational ice spectra show that the \texttt{ENIIGMA} fitting tool can identify both strong and weak absorption features associated with different molecules. In this regard, it will be a useful toolbox for the analysis of future near- and mid-IR observations of YSOs, as for example the data that will be provided by the {\it James Webb Space Telescope}.

\begin{acknowledgements}
We thank the referee, Anthony Jones, for his constructive comments and suggestions that significantly improved this paper. This work benefited from support by the European Research Council (ERC) under the European Union's Horizon 2020 research and innovation program through ERC Consolidator Grant ``S4F'' (grant agreement No~646908) to JKJ. The research of LEK is supported by a research grant (No~19127) from VILLUM FONDEN. WRMR also thanks Dr. Ahmed Gad for the fruitful discussions about the genetic algorithm principles.
\end{acknowledgements}

\bibliographystyle{aa}
\bibliography{References}


\appendix
\section{List of ice laboratory data}
\label{Laboratory_data_list}
A data-set containing 102 laboratory data of astrophysical ices was compiled from different public online databases and are divided into 4 categories, namely, (i) pure and non-processed ices - Table~\ref{ice_list_pure}, (ii) pure and thermally processed ices - Table~\ref{ice_list_pt}, (iii) organic reside samples - Table~\ref{ice_list_orm} and mixed processed ices - Table~\ref{ice_list_mi} and . In the last case, the mixtures are grouped into 3 sub-categories, i.e., non-processed, thermally processed and energetically processed. These ice samples were formed inside a high-vacuum chamber ($P$ $\sim$ $10^{-7}-10^{-10}$~mbar) by the condensation of liquids or gases onto an IR transparent substrate (e.g., ZnSe, CsI) cooled until temperatures around 10~K. In the cases where the ice samples were irradiated by UV, a hydrogen discharge lamp dominated by Ly-$\alpha$ emission with a flux of 10$^{14}$ photons cm$^{-2}$ s$^{-1}$ was used \citep{Schutte2003, Caro2003}. Some samples were irradiated by heavy ions simulating the effect of cosmic rays in the ISM. In this case, different ions were used as a projectile with energies between 0.2$-$632~MeV.

\begin{table}[h]
\caption{\label{ice_list_pure} Laboratory data of pure and non-processed ices used in this paper}
\renewcommand{\arraystretch}{1.0}
\centering 
\begin{tabular}{lccc}
\hline\hline
Label/Temp. & Database& Reference\\
\hline
H$_2$O (15~K) &  NASA Ames   & [1]\\
NH$_3$ (10~K) &  ...   & [2]\\
CH$_4$ (10~K) &  NASA Ames   & [1]\\
CO (12~K) & UNIVAP & [3]\\
CO$_2$ (12~K) & UNIVAP & [3]\\
H$_2$CO (10~K)   & ... & [2]\\
CH$_3$OH (10~K) & NASA Ames  & [1]\\
HCOOH (15~K) &   Leiden DB  & [4]\\
CH$_3$CN (12~K) & UNIVAP  & [3]\\
CH$_3$COOH (12~K) & UNIVAP & [3]\\
CH$_3$CHO (15~K) & Leiden DB&[5]\\
CH$_3$OCH$_3$ (15~K) & Leiden DB &[5]\\
CH$_3$CH$_2$OH (15~K) & Leiden DB   & [5]\\

\hline
\end{tabular}
\tablefoot{[1] \citet{Hudgins1993}, [2] \citet{Gerakines1996}, [3] \citet{Rocha2014}, [4] \citet{Bisschop2007_HCOOH}, [5] \citet{Scheltinga2018}.

}
\end{table}

\begin{table*}
\caption{\label{ice_list_pt} Laboratory data of pure and thermally-processed ices used in this paper}
\renewcommand{\arraystretch}{1.0}
\centering 
\begin{tabular}{lcccc}
\hline\hline
Label & Temperature (K) & Database & Reference\\
\hline
H$_2$O & 40, 75, 100, 120, 140 & NASA Ames& [1]\\
CH$_4$ & 20, 30 &   NASA Ames    &[1]\\
H$_2$CO & 30, 70 &   ...    &[2]\\
CH$_3$OH  & 50, 75, 100, 120 &  NASA Ames    &[1]\\
HCOOH  & 30, 60, 75, 90, 105 &   Leiden DB   &[3]\\
CH$_3$CHO  & 30, 70, 90, 110, 120 &     Leiden DB    &[4]\\
CH$_3$OCH$_3$  & 30, 70, 90, 100 &    Leiden DB      &[4]\\
CH$_3$CH$_2$OH  &  30, 70, 100, 120, 130, 140, 150 &   Leiden DB     &[4]\\

\hline
\end{tabular}
\tablefoot{[1] \citet{Hudgins1993}, [2] \citet{Gerakines1996}, [3] \citet{Bisschop2007_HCOOH}, [4] \citet{Scheltinga2018}.}
\end{table*}

\begin{table*}
\caption{\label{ice_list_orm} Laboratory data of organic residue.}
\renewcommand{\arraystretch}{1.0}
\centering 
\begin{tabular}{lccccc}
\hline\hline
Label/Fraction & ID & Temperature (K) & Database & Reference\\
\hline
HNCO:NH$_3$ (1:1.2) & NH$_4^+$ heating  & 120  &   ...    & [1]\\
H$_2$O:CO$_2$:NH$_3$:O$_2$\tablefootmark{a} (10:2:1.1:1) & NH$_4^+$ heating + UV & 200 & ...  & [1]\\
\hline
\end{tabular}
\tablefoot{\tablefoottext{a}{This data was firstly photolysed by UV and subsequently warmed-up until 200~K.} [1] \citet{Schutte2003}.}
\end{table*}

\begin{table*}
\caption{\label{ice_list_mi} Laboratory data of mixed and processed ices (thermal or energetic) used in this paper.}
\renewcommand{\arraystretch}{1.0}
\centering 
\begin{tabular}{lcccccc}
\hline\hline
\multicolumn{5}{c}{\bf Non-processed mixtures}\\
Label/Fraction & Temperature (K) & Ionizing agent & log$_{10}$(Fluence) & Database  &Reference\\
\hline
H$_2$O:CH$_3$CH$_2$OH (20:1) & 15 & ... & ... &   Leiden DB  &[1] \\
H$_2$O:CH$_3$CHO (20:1) & 16 & ... & ... &  Leiden DB    &[1]\\
CO:CH$_3$OH (1:1) & 15 & ...  & ... &   ...    & [2]\\
\hline
\multicolumn{5}{c}{\bf Thermally-processed mixtures}\\
Label/Fraction & Temperature (K) & Ionizing agent & log$_{10}$(Fluence) & Database  &Reference\\
\hline
H$_2$O:CH$_3$OH:CO:NH$_3$ (100:50:1:1) & 10$-$140\tablefootmark{a} & ... & ... & NASA Ames   & [3]\\
CO:CH$_3$CH$_2$OH (20:1) & 15, 30 & ...  & ... &   Leiden DB     & [1]\\
CO:CH$_3$CHO (20:1) & 15, 30 & ...  & ... &  Leiden DB   & [1]\\
CO:CH$_3$OCH$_3$ (20:1) & 15, 30 & ...  & ... &   Leiden DB      & [1]\\
CO:CH$_3$OH:CH$_3$OCH$_3$ (20:1) & 15, 30 & ...  & ... &    Leiden DB    & [1]\\
CO:CH$_3$OH:CH$_3$CH$_2$OH (20:20:1) & 15, 30 & ...  & ... &    Leiden DB    & [1]\\
\hline
\multicolumn{5}{c}{\bf Energetically processed mixtures}\\
Label/Fraction & Temperature (K) & Ionizing agent & log$_{10}$(Fluence)\tablefootmark{b} & Database &Reference\\
\hline
H$_2$O:NH$_3$ (1:0.5) & 13 & 40 MeV $^{58}$Ni$^{13+}$  & 0, 12, 13.2 & UNIVAP     &[4]\\
H$_2$O:NH$_3$:CO (1:0.6:0.4) & 13 & 46 MeV $^{58}$Ni$^{13+}$  & 0, 12, 13.3 &   UNIVAP    &[4]\\
H$_2$O:CH$_4$ (1:0.6) & 16 & 40 MeV $^{58}$Ni$^{11+}$  & 0, 12, 13 &   UNIVAP     &[4]\\
H$_2$O:CH$_4$ (10:1) & 16 & 40 MeV $^{58}$Ni$^{11+}$  & 0, 12, 13 &   UNIVAP      &[4]\\
H$_2$O:H$_2$CO:CH$_3$OH (100:0.2:0.8) & 15 & 220 MeV $^{16}$O$^{7+}$  & 0, 12, 13 &   UNIVAP  &[4]\\
H$_2$O:HCOOH (1:1) & 15 & 46 MeV $^{58}$Ni$^{11+}$  & 0, 12, 13 &    UNIVAP   &[4]\\
H$_2$O:NH$_3$:c-C$_6$H$_6$ (1:0.3:0.7) & 13 & 632 MeV $^{58}$Ni$^{24+}$  & 0, 12.3, 13.5 & UNIVAP &[4]\\
H$_2$O:CH$_3$OH (1:1) & 15 & 40 MeV $^{58}$Ni$^{11+}$  & 0, 12, 13 &   UNIVAP     &[4]\\
H$_2$O:NH$_3$:CO$_2$:CH$_4$ (10:1:1:1) & 35 & 15.7 MeV $^{16}$O$^{5+}$  & 0, 12, 13 &  UNIVAP     &[4]\\
H$_2$O:NH$_3$:CO$_2$:CH$_4$ (10:1:1:1) & 72 & 15.7 MeV $^{16}$O$^{5+}$  & 0, 12, 13 &   UNIVAP    &[4]\\
CO:NH$_3$ (1:1) & 14 & 5.8 MeV $^{16}$O$^{2+}$  & 0, 12.1, 13 &   UNIVAP      &[5]\\
NH$_3$:CH$_3$OH (1:1) & 14 & 0.2 MeV $^{58}$Ni$^{24+}$  & 0, 12, 13 &  UNIVAP     &[5]\\
H$_2$O:NH$_3$:CH$_3$OH:CO:CO$_2$ (2:1:1:1:1) & 12 & UV (7.3$-$10.5~eV) & 14 & ... & [6]\\
\hline
\end{tabular}
\tablefoot{\tablefoottext{a}{Heating ramp: 10, 40, 80, 100, 120, 140~K.} \tablefoottext{b}{Fluence unit: ions cm$^{-2}$.}  
[1] \citet{Scheltinga2018}, [2] \citet{Cuppen2011}, [3] \citet{Hudgins1993}, [4] \citet{Rocha2017}, [5] \citet{Rocha2019}, [6] \citet{Caro2003}.
}
\end{table*}

\section{Histograms}
\label{Ap_hist}
\begin{figure}
   \centering
   \includegraphics[width=\hsize]{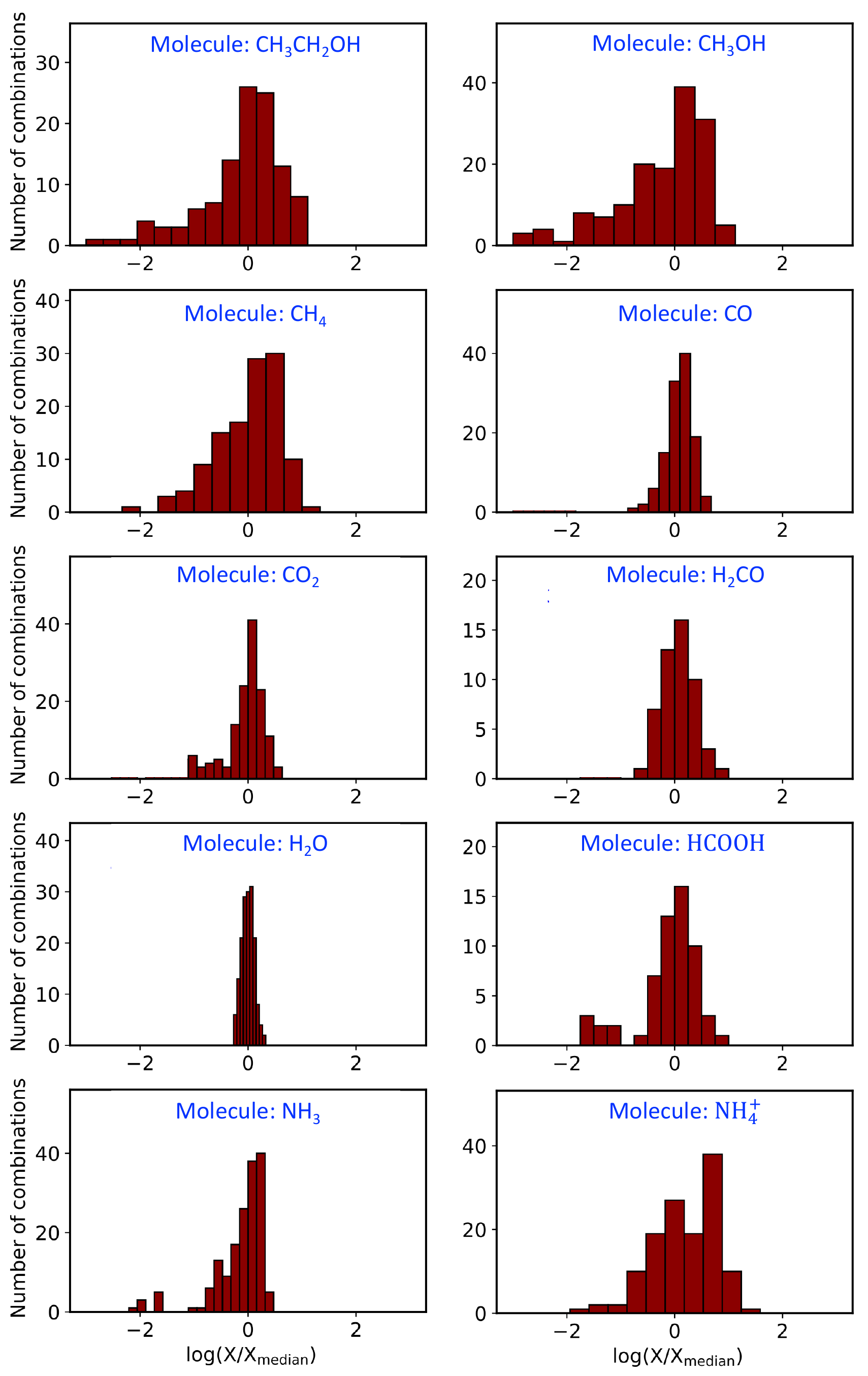}
      \caption{Log-transformed and median centered histograms of the ice column densities derived from all solutions inside 3$\sigma$ confidence interval in the case of Elias~29.
              }
         \label{Hist}
   \end{figure}

All solutions inside a 3$\sigma$ confidence interval out of 1100 combinations (11 components) are analysed with histograms (Figure~\ref{Hist}). The mean, lower and upper limits of the column densities are derived in this analysis (See Table~\ref{ice_cd_el29}). The bin size is proportional to the column density variation and is calculated by the Freedman Diaconis Estimator, that is given by:
\begin{equation}
    h = 2\frac{IQR}{n^{1/3}}
\end{equation}
where $IQR$ is the interquartile range, that is robust to the outliers. In order to reduce the skewness in the distribution, the histograms are log-transformed, although some data still show tails in the distribution. To take into account that skewness factor, the mean and upper limits are calculated by \texttt{scipy.stats.skewnorm} available in the SciPy library\citep{Virtanen2020}. The skewness parameter is calculated, and the statistical quantities derived. The histograms 

\end{document}